\documentclass[usegraphicx]{mn2e}
\newcommand\etal{{\it et~al.}~}
\newcommand\etals{{\it et~al.'s}~}
\newcommand\feh{[Fe/H]}

\begin{document}

\title
[Galactic GCs and SSP Models]
{The Age, Metallicity and Alpha-Element Abundance of Galactic Globular Clusters from Single Stellar
Population Models}

\author
[Mendel, Proctor \& Forbes]
{Jon T. Mendel\thanks{tmendel, rproctor, dforbes@swin.edu.au},
Robert N. Proctor\footnotemark[1]
and Duncan A. Forbes\footnotemark[1] \\
Centre for Astrophysics \& Supercomputing, Mail H39, Swinburne University, Hawthorn, VIC 3122,
Australia}
\maketitle

\begin{abstract} 

Establishing the reliability with which stellar population parameters can be measured is vital to
extragalactic astronomy.  Galactic GCs provide an excellent medium in which to test the consistency
of Single Stellar Population (SSP) models as they should be our best analogue to a homogeneous
(single) stellar population.  Here we present age, metallicity and $\alpha$-element abundance
measurements for 48 Galactic globular clusters (GCs) as determined from integrated spectra using
Lick indices and SSP models from Thomas, Maraston \& Korn, Lee \& Worthey and Vazdekis \etal.  By
comparing our new measurements to independent determinations we are able to assess the ability of
these SSPs to derive consistent results -- a key requirement before application to heterogeneous
stellar populations like galaxies.  

We find that metallicity determinations are extremely robust, showing good agreement for all models
examined here, including a range of enhancement methods.  Ages and $\alpha$-element abundances are
accurate for a subset of our models, with the caveat that the range of these parameters in Galactic
GCs is limited.  We are able to show that the application of published Lick index response functions
to models with fixed abundance ratios allows us to measure reasonable $\alpha$-element abundances
from a variety of models.  We also examine the age-metallicity and [$\alpha$/Fe]-metallicity
relations predicted by SSP models, and characterise the possible effects of varied model horizontal
branch morphology on our overall results.          

\end{abstract} 

\section{Introduction} 

Increased telescope size and improved instrumentation have allowed the observation of ever more
distant objects.  However even with these improvements the vast majority of extragalactic sources
will remain unresolved.  The accurate and reliable analysis of integrated stellar populations is
therefore key to our understanding of formation and evolutionary processes in galaxies.  Through
comparisons of these integrated populations with models of homogeneous stellar systems, or {\it
Single Stellar Populations} (SSPs), recent studies have met with some success in determining ages
and metallicities for both galaxies (e.g. Trager \etal 2000; Terlevich \& Forbes 2002; Maraston
\etal 2003; Proctor \etal 2004a; Maraston 2005; Thomas \etal 2005) and extragalactic globular
clusters (e.g. Forbes \etal 2001; Puzia \etal 2003; Beasley \etal 2005; Pierce \etal 2006).  

Analyses of extragalactic targets have long been dependent on the accurate modelling of stellar
populations.  Empirical approaches to the modelling of integrated light (e.g. Spinrad \& Taylor
1971) have since given way to more rigorous models dependent on a knowledge of underlying physical
processes (i.e. stellar formation and evolution).  These early analyses primarily made use of
broadband colours in deriving their age and metallicity measurements.  However, the limitation of
broadband colours is that they are degenerately sensitive to age and metallicity (i.e. old,
metal-poor and young, metal-rich populations are photometrically identical), heavily restricting the
accuracy of ages and metallicities determined using colours alone.  

The addition of spectral indices, in particular Lick index absorption features (Burstein \etal 1984;
Trager \etal 1998), to stellar population models has afforded the much needed leverage to break this
degeneracy.  SSP models including Lick indices were first assembled by Worthey (1994), who modelled
21 Lick indices and sought to identify those features that were particularly age (e.g. Balmer lines)
or metallicity (e.g. Fe4668, Fe5015, Fe5709 etc.) sensitive and therefore the most useful for
overcoming the observed degeneracy.  More recent works (e.g. Maraston 1998; Vazdekis 1999; Bruzual
\& Charlot 2003; Thomas, Maraston \& Bender 2003; Thomas, Maraston \& Korn 2004; Le Borgne \etal
2004; Maraston 2005; Lee \& Worthey 2005) have focused on including more indices (i.e. the
higher-order Balmer lines H$\delta$ and H$\gamma$) and increasingly complex evolutionary processes
(e.g.  mass-loss and horizontal-branch morphology).

Along these lines, efforts have also been taken to account for known variations in $\alpha$-element
abundance (N, O, Mg, Ca, Na, Ne, S, Si, Ti) with respect to Fe-peak elements (Cr, Mn, Fe, Co, Ni,
Cu, Zn) and their particular effect on Lick index measurements.  Tripicco \& Bell (1995) computed
the effects of variation in C, N, O, Mg, Fe, Ca, Na, Si, Cr and Ti on the 21 Lick indices modelled
by Worthey (1994).  These relative index sensitivities were then used by Trager \etal (2000) to
modify the SSPs of Worthey (1994), facilitating the measurement of ages, metallicities {\it and}
$\alpha$-element abundances for a sample of $\sim$40 elliptical galaxies through a comparison of
H$\beta$, Mg\,$b$ and $<$Fe$>$.  Subsequent calculations of abundance effects have mimicked the work
of Tripicco \& Bell (1995), adding sensitivity calculations for higher-order Balmer lines (e.g.
Houdashelt \etal 2002; Korn, Maraston \& Thomas 2005) and expanding index sensitivities to encompass
a broad range of population metallicities (Korn, Maraston \& Thomas 2005).   

Studies making use of SSPs generally compare modelled Lick line-strengths to those measured from
integrated spectra in order to determine parameters such as age, metallicity and $\alpha$-element
abundance.  To have confidence in the application of these SSPs to observations, it is important
to confirm that they can reproduce independently determined results from colour-magnitude
diagrams (CMDs) and high-resolution stellar spectroscopy.  Globular clusters (GCs) provide a
testbed for SSP models as they represent a coeval and chemically homogeneous stellar
population that should, therefore, be analogous to a synthetic single stellar population.
The availability of resolved observations for Galactic GCs means that accurate ages and
metallicities have already been determined using CMDs, while $\alpha$-element abundances have been
calculated using high-resolution stellar spectra.

Such an analysis was carried out by Proctor \etal (2004b), who fit a sample of 24 Galactic GC
spectra from Cohen \etal (1998) and Puzia \etal (2002) to SSP models from Vazdekis (1999), Bruzual
\& Charlot (2003) and Thomas, Maraston \& Bender (2003) using a multi-index $\chi^2$-minimization
technique, as opposed to 2-dimensional fits e.g. Trager \etal (2000).  They found that it was
possible to recover the known age, metallicity and $\alpha$-element abundance to within $\sim$0.1
dex.  However their small sample size (20 individual GCs) and the relative lack of independent age
and [$\alpha$/Fe] determinations mean a reliable statistical comparison to literature trends was not
possible. 
 
In this study we expand the work of Proctor \etal (2004b), comparing high signal-to-noise
(S/N$\sim$100) spectra of 42 Galactic GCs from Puzia \etal (2002) and Schiavon \etal (2005) to
recent SSP models from Thomas, Maraston \& Korn (2004), Lee \& Worthey (2005) and Vazdekis \etal
(2007).  To these models we apply $\alpha$-element enhancement calculations from both Houdashelt
\etal (2002) and Korn, Maraston \& Thomas (2005).  We are then able to determine ages, metallicities
and $\alpha$-enhancements using the integrated spectra.  A comparison between our determined values
and those from CMD studies (e.g. De Angeli \etal 2005) and resolved stellar spectra (e.g.  Pritzl
\etal 2005) gives an indication as to the reliability of parameters derived solely from integrated
spectral analysis.  

Section \ref{models} contains a brief summary of each of the models, describing their specifics and
the means by which they have been calibrated.  In Section \ref{spectra} we detail the Galactic GC
spectral data used in this work.  This section also contains a discussion of the $\alpha$-element
enhancement models used and an outline of their application.  Section \ref{fitting} describes the
multi-index $\chi^2$ minimization technique that we use for fitting spectra to the SSP models.  In
Section \ref{analysis} we discuss comparisons of parameters we derive from SSPs and literature
values.  Conclusions are given in Section \ref{conclusions}.

\section{Stellar population models} 
\label{models} 

Models from Thomas, Maraston \& Korn (2004; TMK04), Lee \& Worthey (2005; LW05) and Vazdekis \etal
(2007; V07) have been chosen for application to the Galactic GC data.  TMK04 and LW05 models are
both computed using the Worthey \etal (1994) fitting functions and provide Lick/IDS system index
values for a range of ages and metallicities.  Vazdekis \etal provide their models as SEDs, from
which we then measure Lick indices.  We have decided not to include the commonly used models of
Vazdekis (1999) or Bruzual \& Charlot (2003) as they have been studied in a similar fashion by
Proctor \etal (2004b).  

Below is a summary of the essential parameters for each of the SSP models selected for study in this paper.\\

\noindent{\bf Thomas, Maraston \& Korn (2004; TMK04):} These are based on previous work by Thomas,
Maraston \& Bender (2003; TMB03).  Models cover the metallicity range $-$2.25$\leq$[Z/H]$\leq$0.65
with ages from 1 to 15\,Gyrs and are based on isochrones from Cassisi, Castellani \& Castellani
(1997), Bono \etal (1997) and Salasnich \etal (2000).  TMK04 include horizontal branch effects,
providing empirically calibrated Balmer lines modelled for both red and blue horizontal branch
morphologies using the stellar mass loss parameter $\eta$ (Reimers 1975).  Variations in abundance
ratios are tabulated using updated response functions that include the higher-order Balmer lines and
a metallicity dependence as calculated by Korn, Maraston \& Thomas (2005).  These models cover all
25 Lick indices in a wavelength range of $\lambda\lambda$4000--6500\,\AA.  Data can be found at
\verb1http://www.dsg.port.ac.uk/~thomasd/1.\\

\noindent{\bf Lee \& Worthey (2005; LW05):} The Lee \& Worthey models cover a metallicity
range of $-$2.5$\leq$\feh$\leq$0.3 and an age range of 1 to 12\,Gyrs.  Recent Y$^2$ isochrones (Yi
\etal 2001; Kim \etal 2002) are adopted along with post-red giant evloutionary tracks from Yi \etal
(1997).  An additional scaling factor $\eta$ is used to account for stellar mass loss and aids in
matching observed horizontal-branch morphology in Galactic GCs.  SSPs include alpha enhancements of
[$\alpha$/Fe]=0.0,0.3 and 0.6 applied at super-solar metallicities using updated response functions
from Houdashelt \etal (2002).  At sub-solar metallicities, $\alpha$-element abundances are
super-solar and reflect the local abundance-ratio pattern, which includes some metallicity
dependence.  The SSPs model 25 Lick indices from H$\delta_A$ to TiO$_2$, and can be found at
\verb|http://astro.wsu.edu/hclee/wpRGB_all_Lick_2005|.\\

\noindent{\bf Vazdekis \etal (2007;V07):} Models from Vazdekis \etal are based on the previous
models of Vazdekis (1999) and Vazdekis \etal (2003).  These models are presented as SEDs and cover a
metallicity range of $-$2.3$\leq$[Z/H]$\leq$0.2 and ages from 0.1 to 17.5\,Gyrs using Padova group
isochrones from Girardi \etal (2000).  These models are derived using the recent MILES spectral
library (S\'anchez-Bl\'azquez \etal 2006).  Non-solar abundance ratios are not accounted for in
these models, so SSPs represent the local abundance pattern.
SED models are available at \verb|http://www.iac.es/galeria/vazdekis/|.\\

\subsection{SSP Model Calibrations}

For the SSP models outlined above, calibrations have been carried out in order to verify the
accuracy of their index predictions.  This is a key step in the construction of these SSP models, as
the results obtained from their use on extragalactic sources are generally blind (i.e.  there are no
corroborating methods like CMDs or resolved spectroscopy available).  For their calibrations, both
TMK04 and LW05 make use of the P02 Galactic GC data as measured using the Worthey \etal (1994) index
definitions.  The P02 observations were taken with specific care given to their luminosity sampling
in order to obtain accurate cluster spectra with account of stochastic effects.  This careful
sampling means that spectra are representative of the total cluster population and therefore ideal
for the calibration of Lick index models.

Calibrations of the TMK04 models have been well documented in Maraston \etal (2003), TMB03 and
TMK04, which involve assuming an old GC age (12\,Gyrs) and comparing measured GC indices to SSP
predictions.  In Maraston \etal (2003), these comparisons are carried out using index-index
comparison with $<$Fe$>$ (iron-sensitive indices) or Mg\,{\it b} (all other indices) and metallicity
comparison with CMD [Fe/H] determinations (their Figs. 1, 7-11).  Additional evaluations of the
higher-order Balmer lines are carried out through H$\delta$, H$\gamma$ vs. [MgFe] index comparisons
(their Fig. 13).  In all of these   Maraston \etal (2003) note that the GC data of P02 lie as
expected in relation to their SSPs, predicting metallicities and $\alpha$-element abundances
consistent with those from CMD and resolved spectral studies.  These results are reiterated in
TMB03, and the Balmer lines are re-calibrated in TMK04, with the same good agreement being found.
  
Lee \& Worthey (2005) perform similar calibrations for their SSP models, comparing Lick index
measurements of the CBR98 and P02 datasets to their SSP models.  This is done through a comparison
of Lick indices to [Fe/H] as predicted by their SSPs, using metallicities from Harris (1996) for the
GC data.  They, like KMT04, find good agreement between their SSPs and GCs and note that their
models require no zero-point offset to match the GC data.   

At the time of writing, information regarding the calibration of V07 models was unavailable.

\subsection{Non-solar abundance ratios} 
\label{abundance_ratios}

An important consideration in fitting our sample of Galactic GCs is the handling of non-solar
abundance ratios. It is well known that GCs exhibit elemental abundances that differ from those
measured in the Sun (Pilachowski \etal 1983; Gratton 1987), and tabulated response functions have
allowed for these variations to be included in the SSP models.  Tripicco \& Bell (1995; TB95)
modelled response functions for the Lick/IDS index system, providing fractional index variations for
21 Lick indices with respect to 10 elements (C, N, O, Mg, Ca, Na, Si, Ti, Cr and Fe) in three
different stellar types (cool dwarf, main-sequence turnoff dwarf and cool giant).  The TB95
fractional responses were calculated by doubling each element, X$_i$, in turn ([X$_i$/Fe]=+0.3) and
measuring the resultant effects on each index.  While TB95 calculations were carried out using a
5\,Gyr old isochrone, adjusting the relative contribution of their three modelled stellar types
allows the construction of stellar populations with a range of ages, metallicities and
$\alpha$-element ratios.

Work presented by Houdashelt \etal (2002; H02) has sought to update the response functions of TB95.
They have used recent, updated line lists to improve upon the original TB95 calculations and include
the higher-order Balmer lines (H$\delta_{A,F}$ and H$\gamma_{A,F}$) and TiO not modelled by TB95.
H02 have also varied the method by which carbon enhancement is calculated.  Rather than double
carbon (+0.3\,dex) they have chosen to enhance carbon by only +0.15\,dex, seeking to avoid modelling
discrepancies that arise as C/O approaches 1.

Additional work by Korn, Maraston \& Thomas (2005; KMT05) has tested some of the simplifications
made by TB95.  KMT05 confirm the validity of performing all calculations using a 5\,Gyr isochrone by
comparing to the results of calculations made using a 1\,Gyr isochrone.  In this comparison they
find only small deviations between the two results, $\sim$1 percent for G4300 and Fe4348 and
significantly less for all other indices.  In addition, they include calculations for high-order
Balmer lines (H$\delta_{A,F}$, H$\gamma_{A,F}$), TiO and have added metallicity dependence to their
fractional responses, recalculating the same tables as TB95 for six different metallicities from
$-$2.25$\leq$[Z/H]$\leq$+0.67.

In this work we will be using both the KMT05 and H02 fractional sensitivities applied to a variety
of models using the methods described in Trager \etal (2000; T00) and TMB03.  For specific details
of this enhancement application and calibration, we refer the reader to Appendix A.  

For clarity, SSPs that are used as originally published will be referred to by their respective
references (i.e. TMK04, LW05 and V07), while models that we have altered through the use of the H02 and
KMT05 index response functions will be referred to by a combination of the model and enhancement
calculation reference (e.g. TMK+H02, LW+KMT05, V+H02 etc.).

\begin{table}
\centering
\scriptsize
\caption{Number of times each index is clipped for each model set and enhancement method.  Results
for H02 enhanced models are shown, with results for KMT05 enhanced models shown in parentheses.}
\begin{tabular}{lccc}
\hline
Lick 		& & & \\
Index		&LW05	&TMK04 &V07  \\
\hline
H$_{\delta A}$  (\AA)  & 0 (0)   & 0 (0)   & 0 (0)   \\
H$_{\delta F}$  (\AA)  & 0 (0)   & 0 (0)   & 3 (3)   \\
G4300           (\AA)  & 6 (5)   & 9 (8)   & 8 (15)  \\
H$_{\gamma A}$  (\AA)  & 1 (0)   & 0 (0)   & 0 (1)   \\
H$_{\gamma F}$  (\AA)  & 0 (0)   & 1 (1)   & 0 (4)   \\
Fe4383          (\AA)  & 2 (1)   & 1 (1)   & 0 (1)   \\
Ca4455          (\AA)  & 45 (44) & 44 (36) & 0 (1)   \\
Fe4531          (\AA)  & 0 (0)   & 0 (0)   & 0 (0)   \\
C4668           (\AA)  & 33 (38) & 28 (25) & 34 (39) \\
H$_\beta$       (\AA)  & 1 (2)   & 4 (1)   & \ldots  \\
Mg$_1$          (mag)  & 7 (8)   & 5 (6)   & 4 (2)   \\
Mg$_2$          (mag)  & 1 (0)   & 2 (12)  & 1 (3)   \\
Mg~$b$          (\AA)  & 8 (9)   & 14 (23) & 1 (11)  \\
Fe5270          (\AA)  & 2 (1)   & 2 (3)   & 2 (0)   \\
Fe5335          (\AA)  & 0 (1)   & 3 (3)   & 21 (14) \\
Fe5406          (\AA)  & 1 (0)   & 0 (0)   & 2 (2)   \\
Fe5709          (\AA)  & 0 (1)   & 1 (1)   & 1 (2)   \\
Fe5782	        (\AA)  & 20 (24) & 27 (20) & 3 (0)   \\
TiO$_1$	        (mag)  & 1 (1)   & 0 (0)   & 1 (1)   \\
TiO$_2$	        (mag)  & 5 (4)   & 1 (1)   & 2 (2)   \\

\hline
\end{tabular}
\label{cut}
\end{table}

\begin{figure*}
\centering
\includegraphics[scale=0.55,angle=-90]{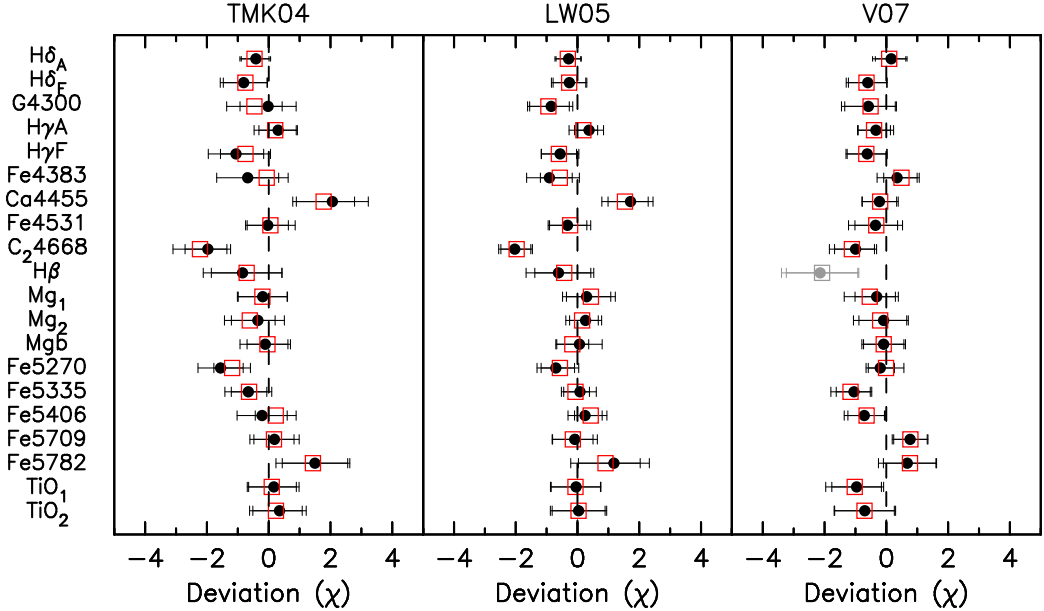}
\caption{Mean deviations of each index for each model set.  Indices clipped in the iterative process
described in Section \ref{fitting} are not included.  Circles and open squares represent fits to
H02 and KMT05 enhanced models respectively.  For V07 based models H$\beta$ was found to be
particularly deviant and excluded from all fits to these models.  It is shown here for comparison
only.}
\label{chis}
\end{figure*}

\section{Galactic GC spectral data} 
\label{spectra}

The Galactic GC spectra used in this study are taken from two different sources.  The first data
are from Schiavon \etal (2005; S05) who provide spectra for 41 GCs chosen to represent a range of
parameters (e.g. age, metallicity, Galactocentric distance, etc.).  These spectra were obtained at
the Cerro Tololo Inter-American Observatory (CTIO) Blanco 4-m telescope with the Ritchey-Chretien
spectrograph and cover a wavelength range of $\lambda\lambda$3350--6430\,\AA~at a resolution of
$\sim$3.1\,\AA~per pixel for central wavelengths.  This allows measurement of all 25 Lick indices
from H$\delta_a$ to TiO$_2$.  For additional details regarding observations, see S05.

Index measurements for the S05 spectra were carried out using index definitions from Worthey \etal
(1994), Worthey \& Ottaviani (1997) and Trager \etal (1998).  Prior to index measurement, spectra
were broadened to the Lick system resolution ($\sim$8-11\,\AA) using a wavelength dependent Gaussian
broadening kernel based on the IDS resolution description given in Worthey \& Ottaviani (1997).
Measurements for 25 indices were produced, however unreliable fluxes around 4546\,\AA~ and
5050\,\AA~ from CCD defects or sky subtraction errors resulted in deviant measurements for the
Fe4531 and Fe5015 indices (see S05 for details).  

The second source of Galactic GC data is from the study of Puzia \etal (2002; P02).  P02 provide
long-slit spectra for 12 Galactic GCs in the wavelength range $\lambda\lambda$3400--7300\,\AA.
Observations were carried out using the European Southern Observatory (ESO) 1.52-m telescope on La
Silla with the Boller \& Chivens Spectrograph with a spectral resolution of $\sim$6.7\,\AA~per
pixel.  Lick line-strengths are given for 25 indices measured and calibrated using both the Trager
\etal (1998) and Worthey \etal (1994) index definitions.  These data will be discussed further in
Section \ref{calibration}.  See P02 for more details regarding observations and line-strength
measurements.

\subsection{Multiple observations} 
\label{multiple} 

The combined dataset used for this analysis includes spectra for 42 unique GCs with 75 observations in
total. P02 contains no duplicate spectra, however S05 include multiple observations and
aperture extractions for several GCs.  In order to assemble a more coherent sample, analyses have
been limited to a single observation for each GC in each study.  In the case of multiple
observations, we have selected to use those that are best fit (i.e. most indices fit with the lowest
$\chi^2$) for the majority of SSP models.  For GCs with multiple aperture extractions (NGC~6284,
NGC~6342, NGC~6441, NGC~6528, NGC~6624 and NGC~7078) we have used the extraction that includes a
wider spatial region than just the FWHM of the slit profile as these, generally, give the best fit.
We find deviations in parameters derived from fits across multiple observations to be small
($\pm$0.015, $\pm$0.031 and $\pm$0.038 in log\,age, [E/Fe] and [Z/H] respectively), and
consequently exclude them from our analysis has little effect on our final results.

\subsection{Calibration to the Lick/IDS system} 
\label{calibration}

The general method for calibrating observations to the Lick/IDS system involves obtaining spectra of
stars in the Lick standard library and using these to calculate line-strength offsets.  Such
calibrations have been carried out for the P02 dataset, however S05 observed Jones library (Jones
1999) standard stars which, owing to the slightly limited spectral coverage of the Jones library
($\lambda\lambda$3820--5410\,\AA~with a gap from 4500\,\AA~to 4780\,\AA), only allow for the
calibration of at most 17 Lick indices.  In an effort to include as many indices as possible in our
SSP model fitting, we have instead chosen to calibrate the S05 dataset using the 11 GCs it shares
with P02.  

Puzia \etal (2002) provide Lick indices measured using both the Worthey \etal (1994; W94) and Trager
\etal (1998; T98) passband definitions.  The differences between W94 and T98 index definitions are
the result of refinements to the wavelength solution of the original Lick/IDS library spectra and
constitute 1.25\,\AA~to 1.75\,\AA~shifts in index definitions.  Central indices (H$\beta$ to Fe5406)
were unaffected by this adjustment as their original definitions were calculated using more finely
calibrated template spectra (Worthey \etal 1994).  

Because of the index adjustment made by T98, their index definitions are the most appropriate for
use on properly wavelength calibrated data; index and pseudo-continuum passbands will fall on the
correct spectral features.  For this reason the ``correct'' index definitions to use for the P02
data are from T98, however concerns have been raised with regards to the calibration of these index
measurements to the Lick/IDS system.  Specifically, P02's W94 measurements were calibrated using
index values published by the Lick group, while their T98 indices were calibrated using indices
re-measured from the published spectra (D. Thomas \& C. Maraston 2006, priv. comm.).  Further
examination of the P02 data shows large offsets between their W94 and T98 measurements, even for
indices whose passband definitions remain the same between W94 and T98.
    
In light of these inconsistencies we have performed calibrations of the S05 data using {\it both}
the W94 and T98 index definitions.  Indices were measured on the S05 spectra using both the W94 and
T98 passband definitions and calibrated using the corresponding data from P02.  This method of Lick
calibration introduces a greater uncertainty in our calculations than if Lick standard stars were
used.  We have therefore adjusted our index errors accordingly, including the rms about the mean
offsets for the common GCs and the rms quoted in P02 for their own calibration to the Lick system
(from their Tables 3 and D1) in our overall error estimates.  The final errors we adopt for this
calibration are shown in Appendix B, Table \ref{app_errors}. 

For the remainder of this work we to show fits to the W94 calibrated data for TMK04 and LW05 based
models to avoid the uncertainties in P02's calibration of their T98 data, discussed above.  All
relevant figures have been reproduced using the T98 calibrated data for comparison and are shown in
Appendix \ref{trager}.    

\subsection{Vazdekis 2007 Models}

Data fit to the V07 models do not require the same Lick/IDS calibrations as data fit to the TMK or LW
based models, since V07 use a well calibrated stellar library.  With this in mind, the S05 data were
broadened using the same wavelength dependent Gaussian discussed above, and indices were measured
using the T98 index definitions.  No additional calibration was performed.

In order to fit the P02 data to the V07 models, we have used the coefficients given for the Lick
calibration of their data measured using T98 definitions (P02's Table 3) to de-calibrate the indices
given in their Table C1.  This results in indices measured using T98 passband definitions on
smoothed, flux calibrated spectra.

In Appendix \ref{ind_ind} we show index-index comparisons for the common GCs in P02 and S05 for the
three difference calibrations discussed above.

\section{Galactic GC fits using SSP models} 
\label{fitting}

\begin{figure*}
\centering
\includegraphics[scale=0.70,angle=-90]{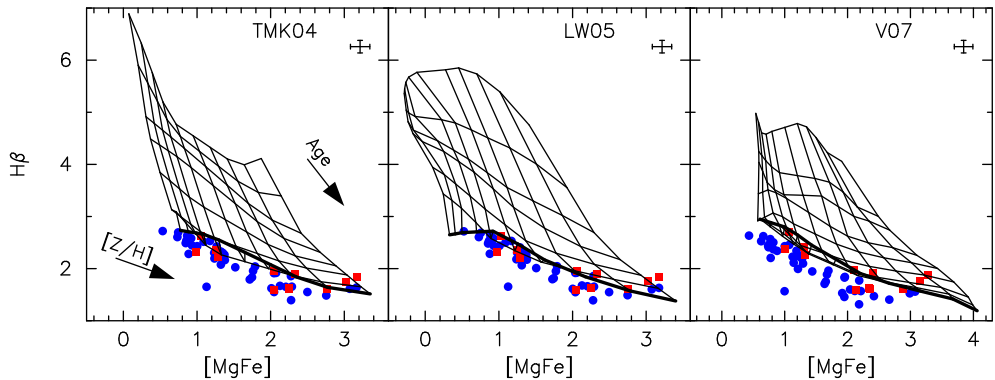}
\caption{SSP grids of H$\beta$ vs. [MgFe] for TMK04, LW05 and V07. Circles (blue) and squares (red)
represent S05 and P02 data respectively.  Grid lines cover metallicities from [Z/H]=0.0 (solar) down
to $-$2.20 in steps of 0.25\,dex.  Age lines are 1, 1.5, 2, 3, 5, 8 and 12\,Gyrs for LW05, with
15\,Gyr lines included for TMK04 and 17\,Gyr lines for V07.  Arrows indicate the direction of
increasing age or metallicity. The average error is shown in the upper right corner (see Section
\ref{calibration}).} 
\label{base_comp} 
\end{figure*}

Having detailed the models and data used, we now turn to a discussion of our fitting technique.  We
have chosen to adopt the $\chi^2$-fitting method discussed by Proctor \etal (2004b), involving the
simultaneous $\chi^2$-minimization of as many indices as possible in order to maximize use of the
available data and break the age-metallicity degeneracy.  This technique has been used previously to
determine ages and metallicities of GCs (Proctor \etal 2004b; Beasley \etal 2005; Pierce \etal 2005;
Pierce \etal 2006), and has been shown to produce more robust results than most individual index
comparisons.  The method involves the $\chi^2$-minimization of measured spectral indices to a grid
of SSP indices corresponding to different metallicities, ages and $\alpha$-element abundances.
Indices that show significant deviations ($\sim$3$\sigma$) from the best fit may be removed and the
fits recalculated.  This process can be continued until no more deviant indices are present and a
stable fit is established.  The fits produced by this method are robust against single deviant
indices and calibration errors, and allow for the reliable identification of trends across multiple
data sets.

The iterative fitting and clipping of indices involved in this multi-index technique makes easy the
identification and omission of indices that are deviant for a majority of the GC spectra.  The NaD
index, for example, is known to suffer heavily from interstellar absorption, and so exhibits large
variations across the data sets when fit; we have therefore excluded this index from all fits. As is
commonly the case in GCs, we find that residuals to best fits of the CN and Ca4227 indices follow a
pattern suggestive of nitrogen enrichment (CN$_1$ and CN$_2$ show positive residuals, while Ca4227
shows a negative residual; e.g. TMB03, Proctor \etal 2004b).  Rather than fit nitrogen as an
independent parameter (e.g. TMB03), we have simply excluded these indices from the fitting
procedure.    

For most of the S05 spectra, measurements of the Fe4531 and Fe5015 indices were inhibited by
``deviant fluxes'' in their index bands, attributed to poorly subtracted sky lines or CCD defects.
Both Fe4531 and Fe5015 have been excluded from all fits to the S05 spectra (See S05 for more
details). 

As discussed by Proctor \etal (2004b) the Fe5015 index showed large deviations between the P02 and
Cohen, Blakeslee \& Ryzhov (1998) data sets, which are possible symptoms of the inconsistencies in
conversion to the Lick/IDS system mentioned in Section \ref{spectra}.  Again, since there is no
absolute way to account for these deviations, the Fe5015 index was excluded from fits to the P02
data.  Taking all of these effects into account, we are left to conduct fits using 20 indices for
the P02 data, and 19 indices for S05.

Table \ref{cut} gives the details of the fits, showing the total number of times that each index was
clipped for a given model set.  Ca4455 was found to be very deviant in the LW05 and TMK04 models and
can be interpreted as a problem in calibrations to the Lick/IDS system, as fits to the V07 models
(un-calibrated data) do not show the same deviations. For all 6 model sets the C4668 index is
particularly deviant, perhaps due in part to its extreme carbon sensitivity (the adopted enhancement
pattern leaves carbon solar scaled).  However the improved fit of this index to V07 models that some
part of this offset could also be due to calibration (as with Ca4455).  We found the H$\beta$ index
to be particularly aberrant in fits to V07 models, and so this index has been removed from all fits
to V07 (both P02 and S05 data).      

\begin{figure*} 
\centering 
\includegraphics[scale=0.55,angle=-90]{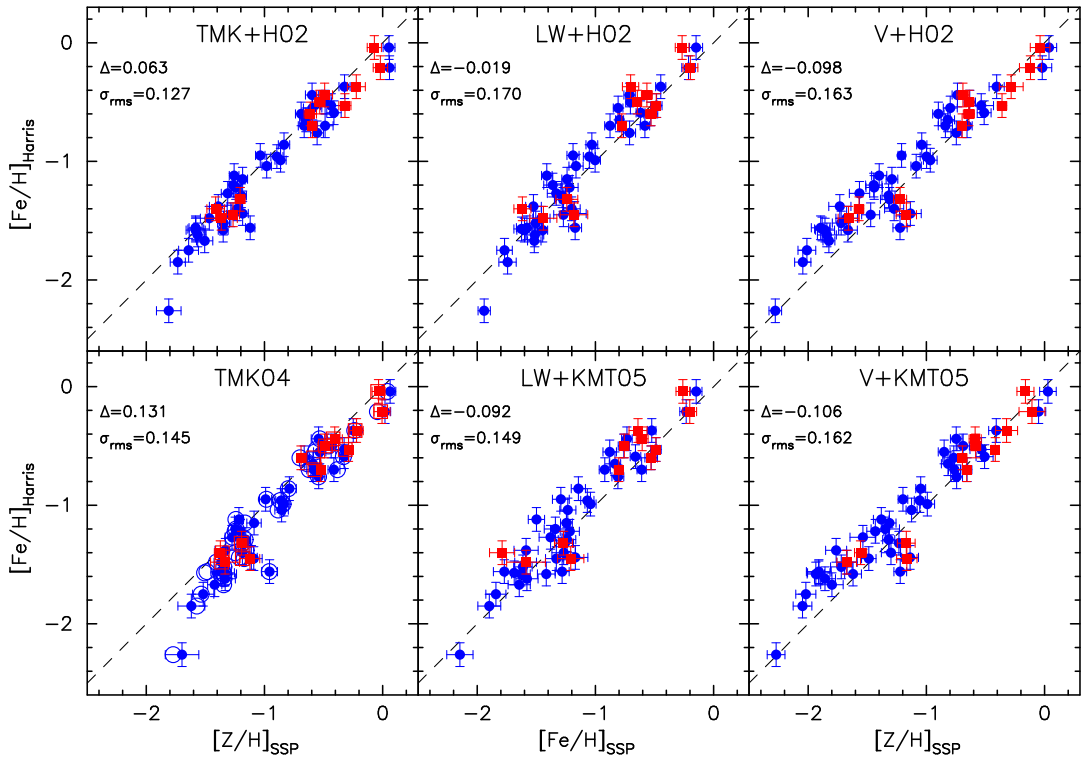}
\caption{[Fe/H] from Harris \etal (1996) plotted against SSP derived metallicities. Numbers in the
upper left corner represent the mean offset from the one-to-one line (dashed) and the $\sigma_{rms}$
scatter about that offset, error bars signify a 1$\sigma$ deviation on our SSP fits and $\pm$
0.1\,dex for the Harris [Fe/H] values.} 
\label{fe_z} 
\end{figure*}

In Figure \ref{chis} we show the mean deviations for our best fits of each index, for each set of
models.  Fits to models using H02 enhancements (TMK+H02, LW+H02 and V+H02) are shown as filled
circles and fits to KMT05 enhanced models (TMK04, LW+KMT05 and V+KMT05) are show as open squares.
Enhancement methods give qualitatively the same fits (to within a fraction of the errors) for a given
model set.  Comparing the quality of fit between LW05 or TMK04 and V07, one sees the benefit of
fitting to models based on a well calibrated stellar library, in this case MILES, reflected in the
reduced deviations seen for most indices.

In all subsequent figures, fits to TMK+KMT05 models are shown as open symbols plotted behind the
parameters derived using TMK04 models, which are shown as filled symbols.  This is done for
comparison only, and analyses are carried using TMK04 models results.

\section{Comparison of SSP Derived Parameters with Literature} 
\label{analysis}

We have measured ages, metallicities and $\alpha$-element abundances for Galactic GCs using several
stellar population models.  In doing this, we are able to compare the SSP derived parameters to
those determined using other methods (i.e. CMDs or resolved stellar spectroscopy) and assess the
validity of SSP determinations.  Establishing the reliability of these SSP predictions is vital as
these models are frequently applied to extragalactic sources (galaxies and GCs) for which
alternative age, metallicity and $\alpha$-element abundance determinations are not available.  It is
important to note that we have therefore conducted fits to age, metallicity and $\alpha$-element
abundance simultaneously, rather than assuming an old age (e.g. Maraston \etal 2003; TMB03), in
order to duplicate the way in which these models are frequently used for extragalactic sources.\\

Fig. \ref{base_comp} shows model grids of H$\beta$ plotted against [MgFe] for TMK04, LW05 and V07.
GC data from S05 and P02 are overplotted.  From Fig. \ref{base_comp}, both TMK04 and LW05 models fit
the data reasonably well, with the largest deviations generally at intermediate metallicities
(--1.0$\leq$[Z/H]$\leq$--0.5) where horizontal branch morphology becomes increasingly uncertain (see
Section \ref{horizontal_branch}).  The comparison of V07 models to data is less encouraging, as the
data lie well below the grids at nearly all metallicities.  This is consistent with our findings
from the $\chi^2$ fits, namely that the H$\beta$ index is particularly deviant when compared to
other indices.

At the low metallicity end, we see that models differ in their coverage of the observed data.
However as it is at these metallicities that stellar libraries become extremely sparse, this
variability is not unexpected.  In all cases data are consistent (within errors) with the models,
which is important to our fitting procedure as stable, accurate fits are difficult to obtain for GCs
whose index values fall outside the range of the models.  

Having briefly looked at the base models, we now turn to a discussion of the ages, metallicities
and $\alpha$-element abundances derived using these SSP models.

\subsection{Metallicity} 
\label{metallicity} 

A direct comparison between TMK04, LW05 and V07 models is not straight-forward, as each have handled
metallicity in a slightly different way.  TMK04 present their models in terms of [Z/H], the total
metallicity, and have accounted for the local stellar pattern in their models (e.g. Wheeler \etal
1989), and so the values of [Z/H] and [E/Fe]\footnote{[E/Fe] is used to describe the measured
enhancement, as it represents an enhancement of N, O, Mg, Ca, Na, Si and Ti as opposed to any single
element.}measured using their SSPs can be used without any adjustment.  In contrast, LW05 models are
supplied as a function of [Fe/H].  As LW05 models do not, as published, account for varying
$\alpha$-element ratios at [Fe/H]$\leq$0, these models carry with them an implicit enhancement,
[$\alpha$/Fe]$_{local}$, equivalent to the local stellar abundance pattern (i.e. [$\alpha$/Fe]=0.3
for [Fe/H]$\leq$-1.0; [$\alpha$/Fe] decreasing from 0.3 to 0.0 as [Fe/H] increases from -1.0 to
solar; [$\alpha$/Fe]=0.0 for [Fe/H]$\geq$0.0).  This pattern must be accounted for in our
measurements, in addition to the enhancement [E/Fe]$_{SSP}$ that we measure from our own enhancement
calculations (see Appendix A).  For LW05 models, [Z/H] is then calculated with Equation A1 using our
measured [Fe/H] and [$\alpha$/Fe]=[$\alpha$/Fe]$_{local}$+[E/Fe]$_{SSP}$. 

\begin{figure*} 
\centering 
\includegraphics[scale=0.55,angle=-90]{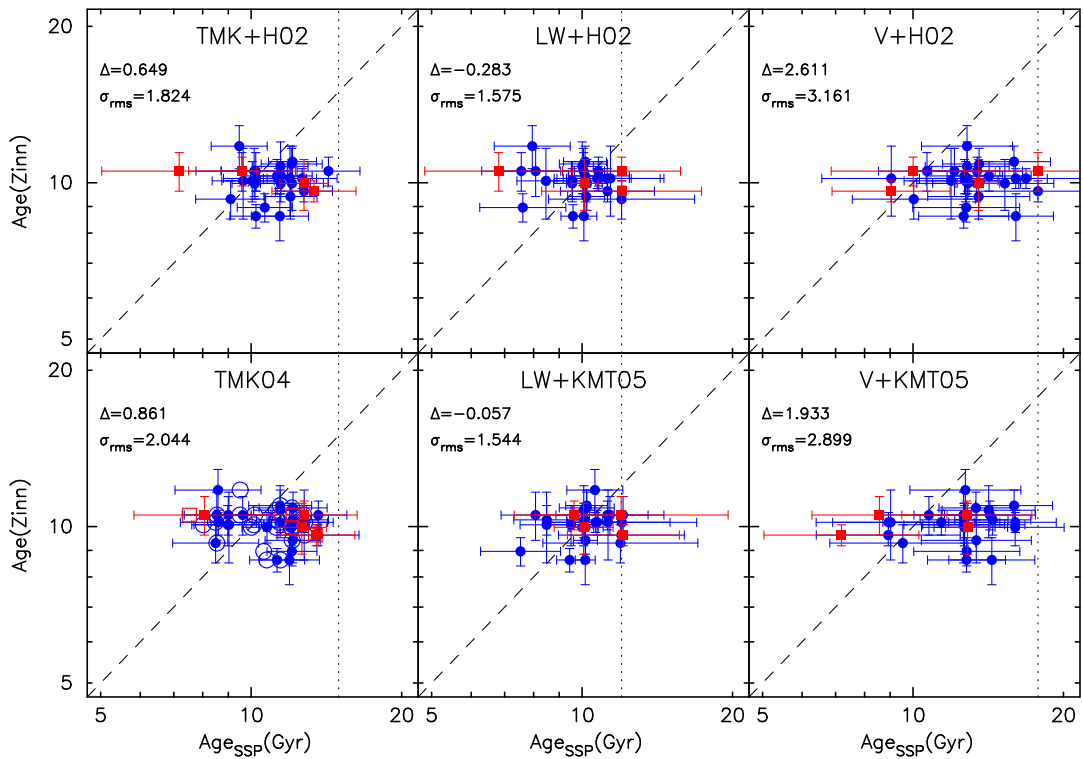} 
\caption{Zinn \& West (1984) scale ages from De Angeli \etal (2005) plotted against SSP derived
ages, the dotted line in this case representing the oldest age in each model set.  Symbols represent
P02 (squares) and S05 (circles).  Numbers in the upper left corner represent the mean offset from
the one-to-one line (dashed) and the $\sigma_{rms}$ scatter about that offset, error bars signify a
1$\sigma$ deviation on our SSP fits.} 
\label{age_age} 
\end{figure*}

The V07 models provide yet another variation, being published as a function of [Z/H], but do not
include varying $\alpha$-element ratio calculations.  Since the [Z/H] we measure is already
includes the afore mentioned local stellar abundance pattern, and our additional enhancement
calculations are applied at constant total metallicity, [Z/H], we can then calculate [Fe/H] in the
same way as LW05, using Equation A1 and [$\alpha$/Fe]=[$\alpha$/Fe]$_{local}$+[E/Fe]$_{SSP}$. 

Compounding these modelling differences is the uncertainty of the Zinn \& West (1984; hereafter
ZW84) GC metallicity scale used in Harris (1996).  While ZW84 is generally quoted as [Fe/H], it is
based on measurements made by Cohen \etal (1983) using the average of the Mg triplet
($\sim$5175\,\AA), 5270\,\AA~and 5206\,\AA~Fe blends.  The output of a particular SSP, then, is
somewhat coloured by assumptions made as to what Harris (1996) {\it actually} represents (be it
iron abundance, [Fe/H], or overall metallicity [Z/H]).  This ambiguity is magnified by evolutionary
tracks and stellar libraries which may or may not have made additional assumptions as to the nature
of ZW84 [Fe/H] values.  

In light of these ambiguities, in Fig. \ref{fe_z} we show [Fe/H] measurements from Harris
(1996) plotted against the most closely related metallicity indicator from each of the models.  For
TMK04 and V07 models, this is [Z/H]$_{SSP}$, however for LW05 this is [Fe/H]$_{SSP}$.  This
represents a fundamental difference in what the models are measuring, and should be kept in mind
when these models are applied to spectra.  This difference is likely due to the several factors
mentioned above, however it does not prevent a qualitative comparison of these models.  In fact all
model variants shown in Fig. \ref{fe_z} find metallicities (either [Z/H]$_SSP$ for TMK04 and V07 or
[Fe/H]$_{SSP}$ for LW05) that are in good agreement with the CMD metallicities from Harris (1996),
regardless of enhancement method.  Looking more closely at the offsets and scatters for each panel
in Fig.  \ref{fe_z} ($\Delta$ and $\sigma_{rms}$ respectively in the upper left corners), the TMK04
and TMK+H02 models give the tightest relations (i.e. lowest $\sigma_{rms}$), albeit with a slightly
larger offset from the one-to-one line than LW05 based models.  V07 models do not seem to follow the
one-to-one line as closely as the other 4 models, generally showing larger offsets and scatters than
either of the other two models.

\subsection{Age} 
\label{age} 

While the range of Galactic GCs ages is quite small ($\sim$2\,Gyrs), a comparison of SSP age
predictions to literature is still useful in evaluating their reliability.  In Fig. \ref{age_age} we
show such a comparison, with CMD determined ages from De Angeli \etal (2005; hereafter D05) plotted
against SSP ages.  Average ages are 10.74$\pm$1.84, 9.38$\pm$1.82 and 11.70$\pm$3.60\,Gyrs for
TMK+H02, LW+H02 and V+H02 models respectively; for models using KMT05 enhancement, mean ages are
10.78$\pm$1.63, 9.60$\pm$1.79 and 11.47$\pm$3.16\,Gyrs.  

Models based on V07 SSPs do a comparatively poor job of GC age prediction, finding mean offsets from
D05 ages and scatters about these offsets ($\Delta$ and $\sigma_{rms}$ in the upper left corner of
each panel in Fig. \ref{age_age}) significantly larger than either TMK04 or LW05 based models.
Modelling uncertainties in the age-sensitive Balmer lines initially seemed the likely culprit for
these large deviations,  however an examination of the H$\delta$ and H$\gamma$ indices did not show
a significant offset (i.e. as is observed in the H$\beta$ index for these models; discussed in
Section \ref{fitting}).  As an additional test to this, fits to V+H02 and V+KMT05 were conducted
with all Balmer lines omitted (H$\delta_{A,F}$, H$\gamma_{A,F}$ and H$\beta$), however no
significant change was observed, i.e. ages were still found to be abnormally high with large
scatter.  As these age deviations are present in {\it many} indices (i.e. more than just the age
sensitive Balmer lines), it seems that either data need some additional calibration to be properly
fit to V07 models, or an additional calibration of the models themselves is needed.

TMK04 and LW05 based models do a good job of reproducing D05 ages, both finding reasonably small mean
offsets ($\Delta$$<$1\,Gyr).  LW+KMT05 models do the best quantitative job of reproducing the CMD
ages of D05, giving both the smallest mean offset ($\Delta$=--0.057) and scatter
($\sigma_{rms}$=1.544), however there are several important caveats to this age analysis. 

Firstly, differing upper age limits for each of the models (15, 12 and 17\,Gyrs for TMK04, LW05 and
V07 respectively; dotted lines in Fig. \ref{age_age}.) likely play some role in the apparent
agreement or disagreement of SSP ages with literature values.  Most notably, fits to LW05 based
models find several GCs with ages equivalent to the upper limit of the models, whereas in fits to
TMK04 derived models, all GC ages are {\it fit} as opposed to being assigned the maximum available
value.

It should also be noted that differences in modelling, especially evolutionary tracks, can affect
the age comparisons shown in Fig. \ref{age_age}.  While Lick indices should not be affected by the
particular set of evolutionary tracks used (Maraston \etal 2003), varied handling of
$\alpha$-element abundance ratios will influence the agreement between SSP and CMD derived ages.
Both De Angeli \etal (2005) and LW05 use $\alpha$-enhanced isochrones (Cassisi \etal 2004 and Kim
\etal (2002) respecively), and so the good agreement in their predicted ages could be a result of
this.  Conversly, both TMK04 and V07 models use solar scaled evolutionary tracks, which have been
shown to produce slightly older age estimates than $\alpha$-enhanced isochrones of a similar
metallicity (Salasnich \etal 2000).  

Additionaly the CMD ages of De Angeli \etal (2005) are subject to uncertainties in their absolute
calibration, being similarly based upon model isochrones.  The ages shown in Fig. \ref{age_age}
should therefore be viewed as measuring the relative agreement of two difference methods of age
measurement, CMD vs. spectral, rather than a comparison of absolute ages.

\begin{figure}
\centering
\includegraphics[scale=0.9,angle=-90]{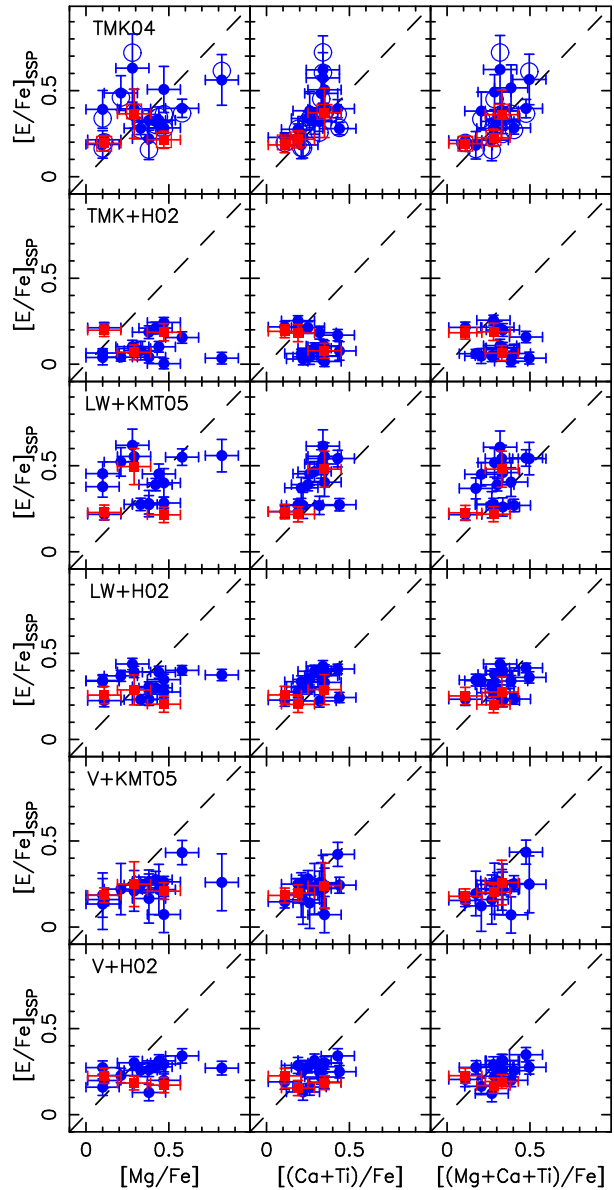}
\caption{SSP derived values of [E/Fe] plotted against high resolution element abundances from Pritzl
\etal (2005). Models and symbols are the same as in Figure \ref{fe_z}.  Error bars signify a
1$\sigma$ deviation on our SSP fits and $\pm$ 0.1\,dex for the high-resolution [$\alpha$/Fe]
values.} 
\label{e_e}
\end{figure}

\subsection{$\alpha$-Element abundance} 
\label{alpha}

The ability of SSP models to accurately measure $\alpha$-element abundances is of great interest as
it can give an indication of formation timescales in galaxies.  Measurements of enhancement using
SSPs, [E/Fe]$_{SSP}$, are shown in Fig. \ref{e_e} plotted against [Mg/Fe],
[(Ca+Ti)/Fe]\footnote{[$\alpha$/Fe] from Pritzl \etal (2005)} and [(Mg+Ca+Ti)/Fe] from Pritzl \etal
(2005).  The agreement between Pritzl \etals [Mg/Fe] and [E/Fe]$_{SSP}$ is poor, likely owing to the
inclusion of additional elements in the SSP enhancement [E/Fe]$_{SSP}$.  Perhaps not surprisingly,
both [(Ca+Ti)/Fe] and [(Mg+Ca+Ti)/Fe] from Pritzl \etal (2005) relate more closely to
[E/Fe]$_{SSP}$.  From Fig. \ref{e_e}, we see that KMT04 and LW+H02 are able to best reproduce the
enhancement values of Pritzl \etal (2005), with [(Mg+Ca+Ti)/Fe] being the best fit while TMK+H02,
V+KMT05 and V+H02 appear to underpredict GC element abundances.
 
We find mean [E/Fe] values of 0.12$\pm$0.08, 0.28$\pm$0.11 and 0.24$\pm$0.06 for TMK+H02, LW+H02 and
V+H02 respectively.  Models using KMT05 enhancement calculations produce higher
[E/Fe]$_{SSP}$=0.28$\pm$0.13, 0.37$\pm$0.12 and 0.22$\pm$012.  All models, with the
exception of TMK+H02, produce mean [E/Fe]$_{SSP}$ values consistent with literature findings of a
constant $\alpha$-element abundance of [E/Fe]$\simeq$0.3 (e.g. Gratton \etal 2004).

\begin{figure*}
\centering
\includegraphics[scale=0.65,angle=-90]{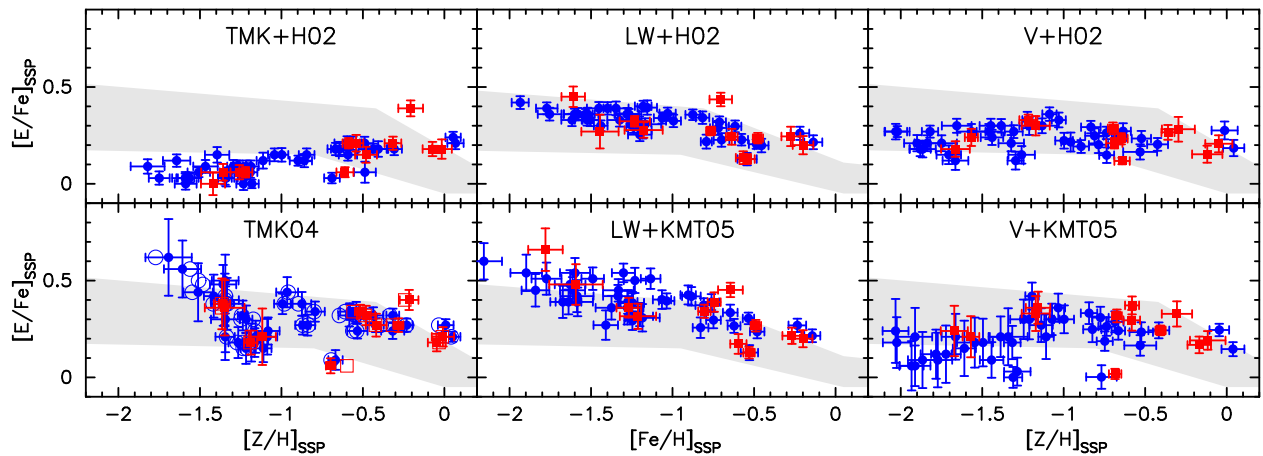}
\caption{Comparison of fitting results for TMK04, LW05 and V07 based models. Symbols are the same as
in previous figures.  The shaded region shows the range of abundances observed in field stars from
Pritzl \etal (2005).  GCs fit at the minimum SSP metallicities are not shown.} 
\label{e_z}
\end{figure*}

To further examine model enhancement predictions, in Fig. \ref{e_z} [E/Fe]$_{SSP}$ is plotted
against the same SSP metallicity indicators as Fig. \ref{fe_z} ([Z/H] for TMK04 and V07, [Fe/H] for
LW05).  The shaded region in Fig. \ref{e_z} represents the range covered by field star data from
Pritzl \etal (2004; their Fig. 4).  This abundance trend with metallicity\footnote{This trend is
generally shown as a comparison with [Fe/H] , however here we have plotted [E/Fe]$_{SSP}$ against
the independently determined metallicity measure for each model.} is generally attributed to the
increased influence of Type Ia SNe at later times when higher metallicity stars formed, and so it is
not surprising that GCs formed at a similar epoch (i.e. metallicity) are found to follow this same
$\alpha$-enhancement pattern.  At higher metallicities, i.e. [Fe/H]$>$--0.5, evidence for GCs
exhibiting the same $\alpha$-enhancement ``down-turn'' as field stars is less certain, however this
is largely due to the small number of high-metallicity GCs relative to lower metallicities.

With regards to the SSPs fit here, TMK04, LW+H02 and V+H02 most closely match the observed field
star abundance pattern.  LW+KMT05 models seem to over-predict enhancements at all metallicites (as
evident from the mean [E/Fe]$_{SSP}$=0.37), while TMK+H02 greatly {\it under}-predict enhancement at
low metallicities and are inconsistent with the field star pattern.  All KMT05 enhanced models show
some deviation at low metallicities, either over-predicting (i.e. TMK04 and LW+KMT05) or
under-predicting (i.e. V+KMT05) enhancement.  The variation seen in Fig. \ref{e_z} between TMK04 and
LW+KMT05, which find higher [E/Fe]$_{SSP}$ at low metallicities, and V+KMT05, which find lower
[E/Fe]$_{SSP}$ at low metallicities is a result of the V07 models being offset from the data (as
seen in Fig. \ref{base_comp}).  These deviations, however, are accompanied by increasing error
distributions and so are roughly consistent with the bulk of the data.  This will be discussed in
more detail below.

\subsubsection{Houdashelt \etal (2002) vs. Korn, Maraston \& Thomas (2005) Enhancement}

As previously mentioned, one marked difference between the KMT05 and H02 enhanced model sets is the
tendency for KMT05 enhanced models to show odd enhancement behavior at low metallicities, be it the
increased [E/Fe] values at low metallicities in the TMK04 or LW+KMT05 models or the abnormally low
[E/Fe] values in V+KMT05 models.   
 
The primary difference between the KMT05 and H02 enhancement calculations is the inclusion of
metallicity dependent index sensitivities by KMT05.  In their calculations, KMT05 found that indices
at low metallicities are relatively insensitive to variations in $\alpha$-element abundance.  This
insensitivity means that all of the SSP grid lines ``pinch'' together at low metallicities, resulting in
indices that may be only slightly enhanced in line-strength relative to [E/Fe]=0.0 being measured as
having elevated enhancement (as is the case in both TMK04 and LW+KMT05 models) and increased
error distributions (e.g. V+KMT05). 

To test that the primary difference between H02 and KMT05 enhanced is, in fact, the added
metallicity dependence rather than an overall shift in enhancement calculation between H02 and KMT05
we have constructed two new sets of models (using LW05 and TMK04) using the [Z/H]=0.0 index
sensitivities from KMT05 applied at all metallicities, making them comparable to H02 enhanced
models.  When we compare ages, metallicities and [E/Fe] values between these models and their H02
enhanced counterparts (TMK+H02 and LW+H02) we find that differences are of order an interpolation
step ($\pm$0.025\,dex in log(age) and metallicity, $\pm$0.03\,dex in [E/Fe]).  Deviations between
KMT05 and H02 models, then, are almost entirely due to the metallicity dependence added by KMT05.

\subsection{Age-Metallicity Relation} 
\label{met_age}

The age-metallicity relation (AMR) for Galactic GCs is well established and shows that Galactic GCs
are generally old (e.g. Salaris \& Weiss 2002; Beasley \etal 2005; Puzia \etal 2005; De Angeli \etal
2005).  De Angeli \etal (2005) have most recently examined the AMR of Galactic GCs using {\it HST}
imaging and found that very low metallicity GCs ([Fe/H]$\leq$--1.4) are old ($\sim$11\,Gyrs) with
very low scatter in their ages (0.06\,Gyrs), however at intermediate metallicities GC ages show
considerably more variety in their ages, ranging from 7.5 to 11\,Gyrs.  

\begin{figure*}
\centering
\includegraphics[scale=0.55,angle=-90]{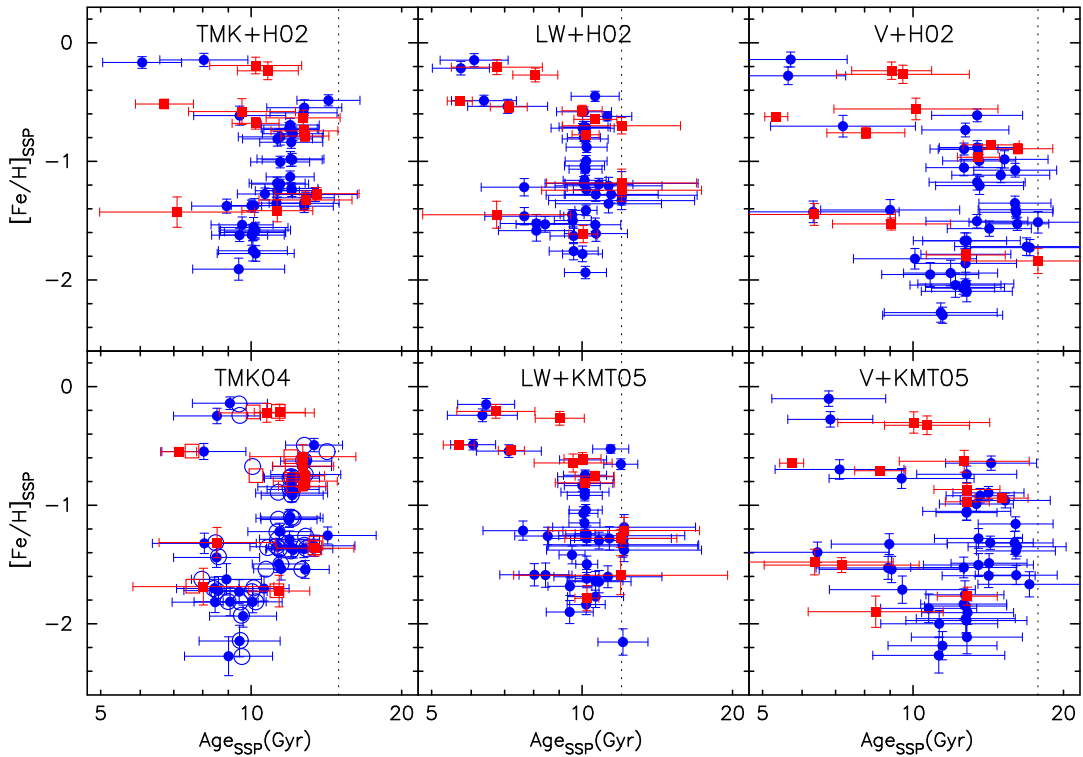}
\caption{[Fe/H] vs. age as derived from our SSP fitting.  Symbols and models are the same as in
previous figures.  The dotted line represents the maximum age for a particular SSP.} 
\label{amr}
\end{figure*}

While the precision of SSP age measurements is not fine enough to delineate between slight
variations in age (our general uncertainty is $\sim$2\,Gyrs), it is of interest to test whether SSP
models can reproduce the overall trend of uniform, old GC ages.  In Fig. \ref{amr} we plot the AMR
as derived from each of the SSP models.  

Models based on TMK04 produce AMRs that reveal an odd trend of increasing age towards higher
metallicity.  It should be noted, however, that age determinations at the lowest metallicities are
highly uncertain due to possible variations in the horizontal branch morphology.  As a result, while
there is a suggestion of a positive AMR slope, the data are consistent with a uniformly old GC
system.  TMK04 especially does an excellent job of reproducing a generally old GC population,
finding only one GC younger than 8\,Gyrs.

LW+H02 and LW+KMT05 model fits show GCs consistently old at metallicities [Fe/H]$<$--1, however
they have a high metallicity ``tail'' towards younger ages.  CMD age determinations for these
highest metallicity GCs, NGC\,6528 and NGC\,6553, suggest that their actual ages are old
(11-13\,Gyrs, Zoccali \etal 2001; Feltzing \& Johnson 2002) and consistent with the rest of the
Galactic GC system.  Comparing LW05 based models with TMK04, we note that TMK04 does not appear to
have the same problem at high metallicities, finding ages for NGC\,6528 and NGC\,6553 consistent
with CMD determinations.   Errant age measures from LW05 based models could be indicative of issues
with this particular model set at higher metallicities, but we are unable to comment in more detail
on the quality of fits at higher metallicites due to a lack of data points.  At metallicities
[Fe/H]$\leq$--0.4 the LW+H02 model is in excellent agreement with CMD based AMRs (e.g. De Angeli
\etal 2005). 

V07 based models have difficulty in producing an AMR consistent with the known GC AMR (e.g. De
Angeli \etal 2005).  This is almost certainly due to the age determination problems discussed
earlier (see Section \ref{age}), which makes any useful AMR determination nearly impossible.  Again,
the large age variations seen when fitting to these models ($\sim$10\,Gyrs) may be alleviated via
calibration to the MILES stellar library. 

\subsection{Horizontal Branch Morphology} 
\label{horizontal_branch} 

Balmer line indices (H$\delta_{A,F}$, H$\gamma_{A,F}$ and H$\beta$) are particularly sensitive to
the presence of hot stars, becoming weaker as temperatures decrease.  This lends to their use as age
indicators as the decrease in main-sequence turnoff luminosity, and therefore temperature,
associated with an aging stellar population is echoed strongly in the measured Balmer line
strengths.  However as older populations are considered ($>$10\,Gyr), the increased presence of hot
horizontal branch (HB) stars causes Balmer line strengths to increase, leading to an age
degeneracy at low metallicities, with very old stellar populations appearing young. 
 
The modelling of these HB morphologies is particularly difficult, as the interplay of contributing
effects (e.g. mass-loss, metallicity, dynamical effects etc.) is not known well enough to be
modelled in detail (i.e. based purely on theory).  Modelling varying HB morphologies, then, has been
done primarily via prescriptive methods.  Maraston \& Thomas (2000) find that a mass-loss
prescription is able to reproduce the strong Balmer lines found in old elliptical galaxy populations, as
well as the trends of increasing H$\beta$ line strengths in low-metallicity Galactic GCs.

\begin{figure}
\centering
\includegraphics[scale=0.42,angle=-90]{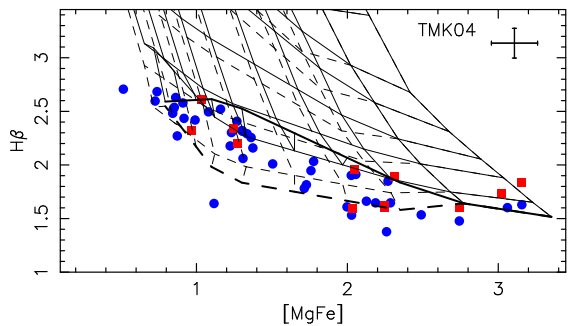}
\caption{TMK04 SSP grids for blue (solid lines) and red (dashed lines) horizontal branch
morphologies.  P02 (circles) and S05 (squares) GC data overplotted.}
\label{rhb_vs_bhb}
\end{figure}

Of the SSP models used here, only TMK04 allow for a variation of HB mophology in their models,
supplying two sets of empirically calibrated Balmer line indices (H$\delta_{A,F}$, H$\gamma_{A,F}$
and H$\beta$), one each for Red and Blue HB morphology (see Maraston \& Thomas 2000 for details).
In Fig. \ref{rhb_vs_bhb} we show a comparison of the BHB (solid lines) and RHB (dashed lines) grids
supplied by TMK04.  This figure illustrates the need for a consideration of variable HB morphology,
especially at intermediate metallicities where hot horizontal branch stars begin to cause large
variations in Balmer line strengths.    

\begin{figure}
\centering
\includegraphics[scale=0.52,angle=-90]{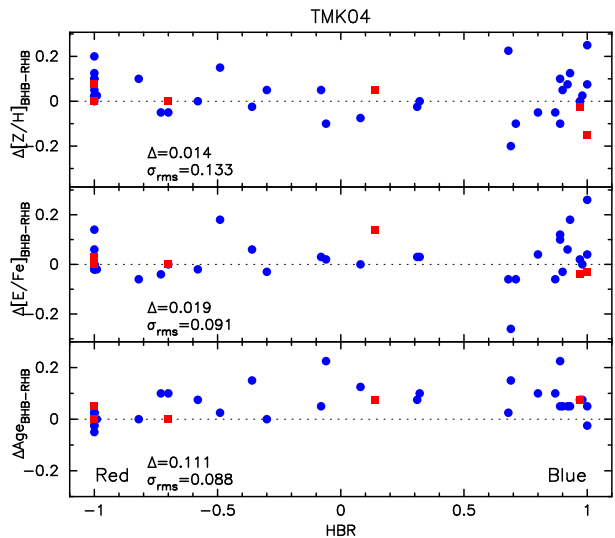}
\caption{Difference in metallicity, enhancement and age between red and blue horizontal branch
models plotted against horizontal branch ratio (HBR) from Harris (1996).  Symbols are the same as
previous figures.  Offsets and scatters are shown in the lower left corners for all GCs with
HBRs$\leq$0.9.}
\label{hb_comp}
\end{figure}

In an effort to better quantify the HB effects in our results, we have performed a second set of
fits to the RHB models of TMK04 using the same techniques described in Section. \ref{fitting}.  In
Fig. \ref{hb_comp} we show the difference between age, metallicity and $\alpha$-element abundance
derived using BHB and RHB models plotted against the GC horizontal branch ratio (HBR) from Harris
1996 and Zoccali \etal (2000).  Since the HBR is based purely on numbers of stars in a given
branch\footnote{HBR=$\frac{B-R}{B+V+R}$} it is an excellent, independent means of determining HB
morphology.

The first thing to note in Fig. \ref{hb_comp} is that GCs with HBRs$>$0.9 are almost entirely blue,
and so have been excluded from the statistics shown in the bottom left of each panel.  As the RHB
modelling in TMK04 is limited to Balmer line indices, it is not surprising that the changes seen in
[Z/H] and [E/Fe] as a result of this modelling are small.  In particular, the offsets for both are
dominated by the scatter.  Ages are most strongly affected by varying HB morphology in the TMK04 SSP
models, however even mean offset for these ($\sim$0.11) is of order the error for our age
determinations ($\sim$0.1).  

While the results of this comparison do show that HB morphology is important to individual indices
(e.g. H$\beta$ in Fig. \ref{rhb_vs_bhb}), it does no appear that it greatly affects the properties
that we derive using the multi-index fitting technique.  Most importantly, the offsets that we see
as a result of varying the HB modelling are not large enough to significantly change the results of
our analyses.

\section{Conclusions}
\label{conclusions}

Using integrated spectra of Galactic GCs, we have tested SSP model predictions of age,
metallicity and $\alpha$-element abundances using Lick indices.  The multi-index
$\chi^2$-minimization technique used has allowed us to measure GC stellar population parameters
consistent with published values, even in situations where data are poorly fit in a single
index-index space.

Although Galactic GCs cover a wide range of metallicities (--2.28$\leq$[Fe/H]$\leq$--0.04 in our
sample), our GC sample only represent a limited parameter space in age and $\alpha$-element
abundance (the mean age of our GCs is 10.23$\pm$0.83\,Gyrs from De Angeli \etal 2005, and
[$\alpha$/Fe]=0.30$\pm$0.09\,dex from Pritzl \etal 2005).  To expand this work, SSP model parameter
space should be explored further by obtaining integrated spectra for Galactic GCs covering a broader
range of ages (e.g. Terzan 7 or Pal 12 with intermediate ages) and $\alpha$-element adundances (e.g.
Rup 106 or Pal 12 with sub-solar $\alpha$-element abundances).

We find metallicity to be the most robust parameter, showing almost no sensitivity to the different
enhancement calculations.  We note that differences in the construction of models have lead to a
fundamental difference in the metallicities they predict.  In particular, [Z/H] measurements from
TMK04 and V07 models are mostly closely related to CMD [Fe/H] from Harris \etal (1996).  However for
LW05 models, SSP predictions for [Fe/H] most closely match the metallicities of Harris \etal (1996).
This does not affect our conclusions, however is an important caveat to keep in mind when appling
these models.  

Age determinations using TMK04 and LW05 based models are very reliable for the old GCs
of the Milky Way.  However, V07 models have difficulty in recovering reliable age measurements,
giving GCs that are too old and with a very large scatter.
 
Of the models tested here, only TMK04 models provide measurements of $\alpha$-element abundances at
all metallicities.  However, we have shown that with a relatively simplistic application of Lick
index sensitivity calculations from either Korn, Maraston \& Thomas (2005) or Houdashelt \etal
(2002) we are able to recover reasonable $\alpha$-element abundances (i.e. consistent with the
literature) from each SSP model.   

We have shown that HB morphology is an important consideration, as it can dramatically affect the
age sensitive Balmer line indicies.  However, in spite of these Balmer line variations, we find
changes in SSP determined ages as a result of varying HB morphology (i.e. blue or red HB models) are
relatively small for the majority of GCs when using the multi-index $\chi^2$ fitting method.
Determinations of metallicity and $\alpha$-element enhancement are relatively robust to changes in
HB morphology, a key result for extragalactic GC studies where direct determinations of HB
morphology are unavailable.

\section*{Acknowledgements}

The authors thank Mike Beasley for his ongoing comments as well as Thomas Puzia, Claudia Maraston,
Ricardo Schiavon, Scott Trager and Guy Worthey for their comments and conversations.  We would also
like to thank the referee Daniel Thomas for his helpful suggestions.  We acknowledge the analysis
facilities provided by IRAF, which is distributed by the National Optical Astronomy Observatories,
which is operated by the Association of Universities for Research in Astronomy, Inc., under
cooperative agreement with the National Science Foundation.  We also thank that Australian Research
Council for funding that supported this work.

\clearpage
\clearpage
\clearpage
\clearpage

\begin{appendix}
\label{detail}
\section{Details of Enhancement Application}

Here we provide some more details of our ad hoc enhancement applications involving the calculations
of H02 and TMK05.  In summary, these methods involve the adjustment of $\alpha$-element abundances
at a fixed metallicity. Using the relation 
     
\begin{equation}
[Z/H]=[Fe/H]+A[\alpha/Fe]
\label{zrelate}
\end{equation}

\noindent which, applied at constant metallicity leads to the following equation from (T00):

\begin{equation}
A=-\frac{\Delta[Fe/H]}{\Delta[\alpha/Fe]}
\label{calcA}
\end{equation}

\noindent where A varies depending on which elements are selected as enhanced or depressed. In this
work we consider N, O, Mg, Ca, Na, Si and Ti as $\alpha$-group elements, while Cr and Fe are
depressed.  We leave C unchanged (i.e. solar-scaled).  This enhancement scheme mimics the work of
TMK04, allowing us to compare our TMK+KMT05 models to published TMK04 models that use these
same index sensitivity calculations.  Using Equation \ref{calcA}, we find A=0.934 when C is excluded
from the enhanced group and solar-scaled.  Following TMB03, the enhancement applied using the
following equation:    

\begin{equation}
\label{ind_calc}
I_{new}=I_{ssp}\prod^n_{i=1}\exp[R_{0.3}(i)]^{(\Delta[X_i]/0.3)}     
\end{equation}

\noindent where the quantity $R_{0.3}(i)$\footnote{$R_{0.3}(i)=\frac{1}{I_0}\frac{\partial
I}{\partial[X_i]}0.3$} refers to the index change resultant from increasing the abundance of the
i$^{th}$ element by 0.3\,dex (adopting notation from T00).  This equation is arrived at by assuming
ln\,$I$ is a linear function of [$X_i$] and using a Taylor expansion to approximate the effects of
abundance ratio changes (see TMB03 for details).\\

\subsection{Negative indices}
\label{negatives}

One of the problems that arises when applying abundance ratio adjustments is that, in some cases,
Lick indices have negative values.  In applying the expansion of ln\,$I$ (Equation \ref{ind_calc})
however, it is implicit that the index values cannot be negative, and in fact asymptotically
approach zero at low abundances (TMB03; T00).  KMT05 handle this problem by applying their computed
index response directly to the flux of their absorption lines, yielding a positive result.  Lacking
the means to apply corrections in this manner, we have adopted the method used by TMB03, in which
the lowest value of a particular index at a given age is taken as the zero point, and all other
indices are scaled to reflect this zero point shift.  In the case of the high-order Balmer lines
this occurs at the highest metallicity, while for most other Lick indices it occurs at the lowest.
For C4668 and NaD, the low points do not occur at either end of the metallicity scale, indicating an
inflection point in the indices at an intermediate metallicity.  For these indices, values from the
local minima are adopted. The adjustment, using notation of TMB03, is defined as follows:

\begin{equation}
\label{delta}
\delta \equiv I_{low}-|I_{low}|
\end{equation}

\begin{figure}
\centering
\includegraphics[scale=0.43,angle=0]{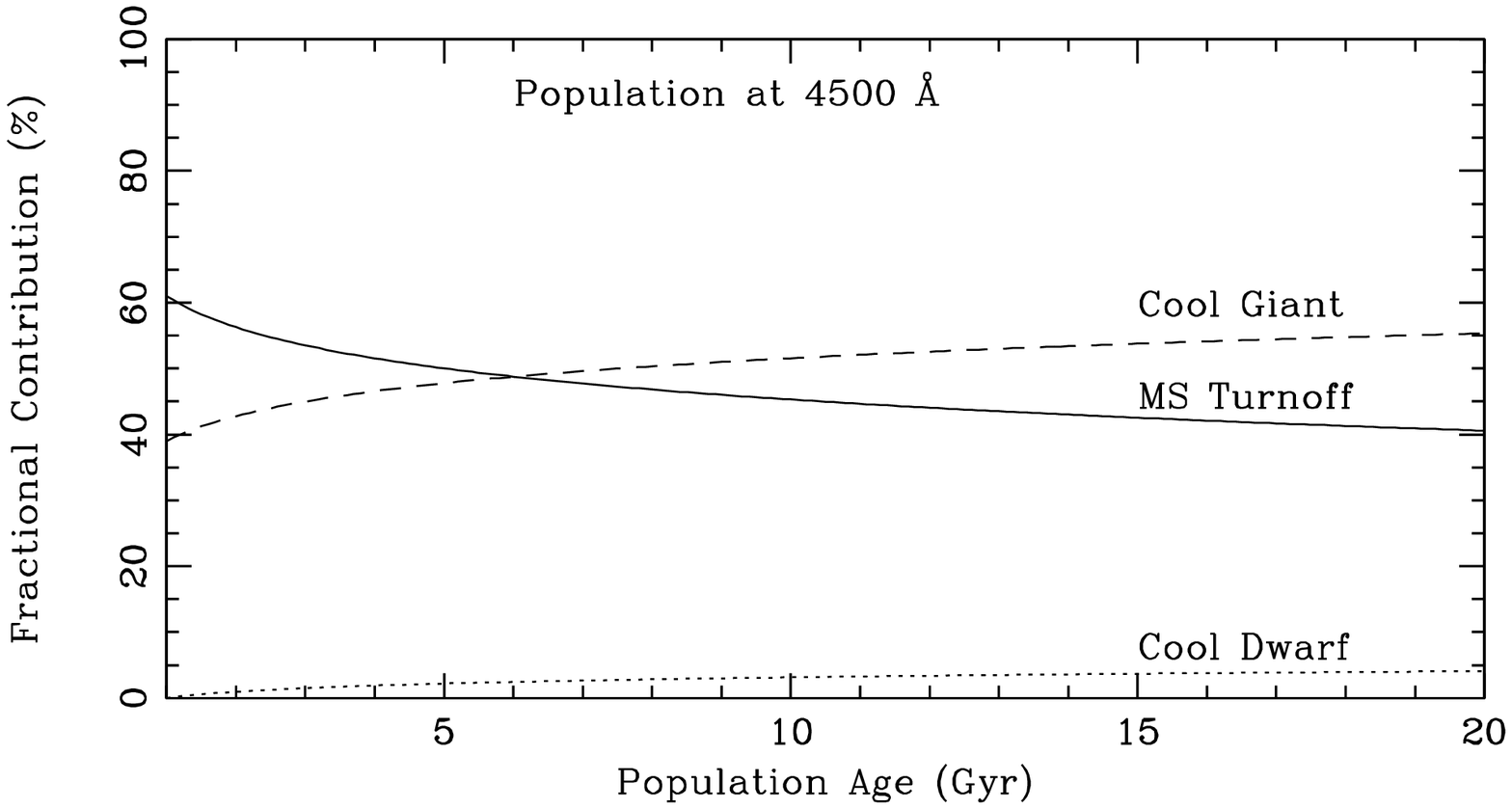}
\caption{Our adopted contributions of each stellar type with age as approximated using Fig. 41 of
Worthey (1994).  This allows for the luminosity of older populations to be increasingly affected by
stars occupying the giant phase of evolution, and proved important in matching the behaviour of the
TMK04 models.}
\label{age_pop}
\end{figure}

This process is summarized in equation 9 from TMB03:

\begin{equation}
\label{ind_adj}
I_{new}-\delta=(I_{ssp}-\delta)\prod^n_{i=1}\exp\left(\frac{1}{I_0-\delta}\frac{\partial I}{\partial
[X_i]}0.3\right)^{(\Delta[X_i]/0.3)}     
\end{equation}

\noindent where $\delta$ is now the zero-point index adjustment, and I$_0$ is the absolute index
value taken from the index response tables.  The key difference between TMB03's and our application
of this zero-point adjustment is that TMB03 adjust each evolutionary phase individually.  Here we
are left to apply this adjustment to the integrated index measurements.

\subsection{Additional Considerations}
\label{application}

In calculating our fractional index sensitivities, we have included several additional effects not
discussed in Sections \ref{abundance_ratios} and \ref{negatives}.  Firstly, the fractional responses
of KMT05 have been adopted strictly in a differential sense, using the SSP index values ($I_{ssp}$) in
place of their I$_0$ values.  While shifting from I$_0$ to I$_{ssp}$ should generally improve
accuracy, for indices in which the lowest index value is very small (i.e.
$I_{ssp}-\delta\leq$\,0.05) the calculated fractional responses are overestimated, since
$R_{0.3}(i)\propto (I_{ssp}-\delta)^{-1}$.  To account for this we have applied a correction to
Equation \ref{delta} such that

\begin{equation}
\label{delta_corr}
\delta \equiv I_{low}-|I_{low}|-1
\end{equation}

\noindent This adjustment increases all indices relative to the zero point, and should have little
effect on the applied fractional enhancements except in the cases of very low index values.

\begin{figure}
\centering
\includegraphics[scale=0.43,angle=0]{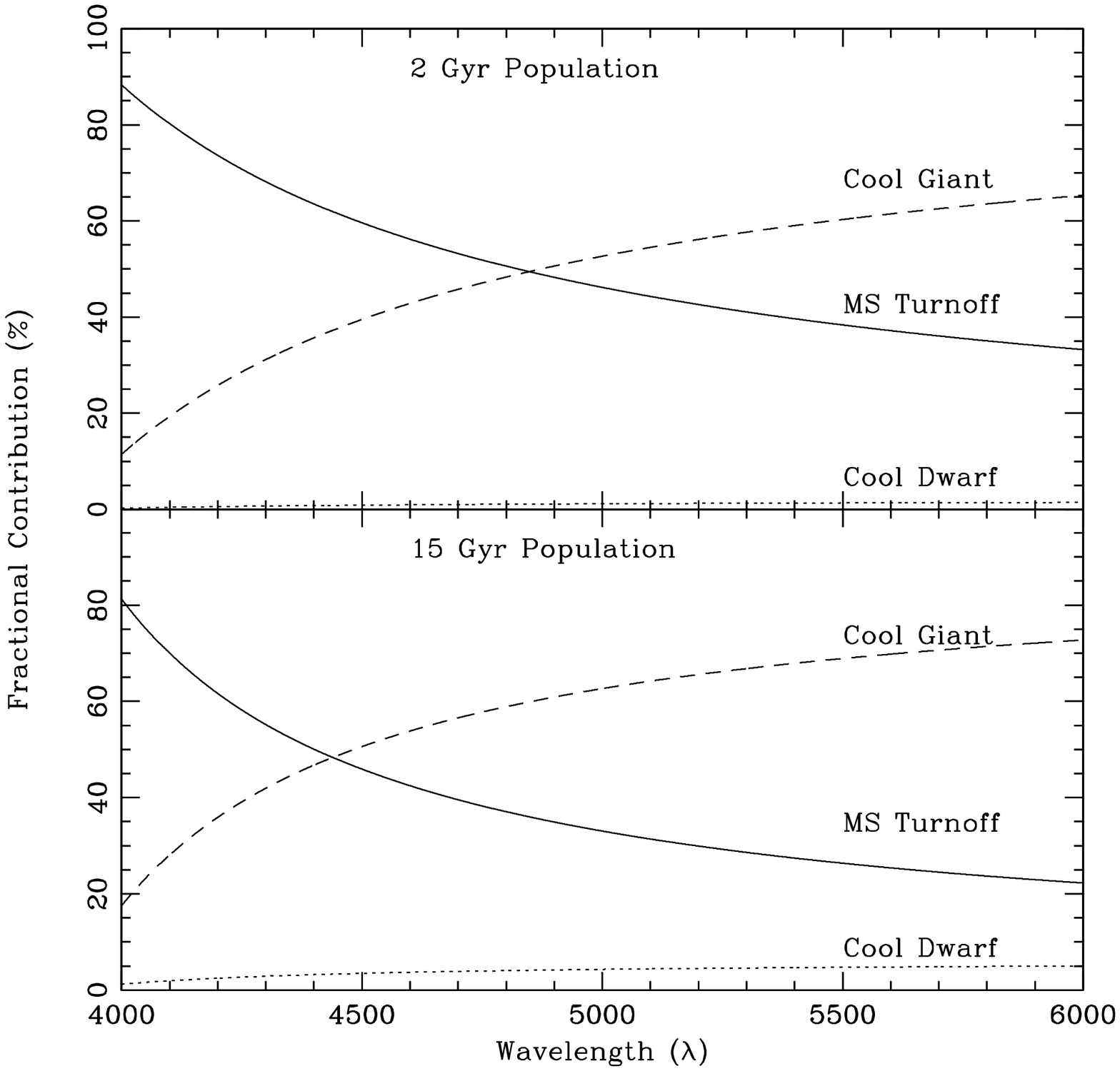}
\caption{Our adopted contributions of each stellar type as they varies with wavelength.  These show
the same trends as noted by Worthey (1994), that for younger populations the equivalence point for
giant and main sequence stars shifts to higher wavelengths.}
\label{lambda_pop}
\end{figure}

Secondly, additional considerations were included to account for variations in the flux contribution
of a given evolutionary type with respect to both age and wavelength.  This was done using estimated
values from Fig. 41 from Worthey \etal (1994).  The results of our approximations are shown in Figs.
\ref{age_pop} and \ref{lambda_pop}.  The effect of this added age and wavelength dependence is
generally 5-10 percent, however can be as much as 50 percent for the bluest indices in the youngest
populations. We stress that these are approximations only, made to include some handling of varying
flux contributions for different populations.  

Since the published TMK04 models make use of the KMT05 enhancement calculations, our goal in
applying enhancements was to mimic the TMK04 models with our own TMK+KMT05 calculations.  To this
end, we have compared the TMK+KMT05 models to the TMK04 models for twice solar ([E/Fe]=0.3)
grids.  The results of this comparison are shown in Columns 2 \& 3 ([Z/H]$<$0 and [Z/H]$\leq$0
respectively) of Table \ref{conv_errors}.  We find the agreement to be acceptable, particularly at
the lower metallicities where the majority of our GCs lie.  In nearly all cases deviations we find
are less than our adopted errors for conversion to the Lick/IDS system and should, therefore, have
negligible effects on parameters (age, metallicity and $\alpha$-enhancement) derived from fits.  The
C4668 index is an exception to this, showing significant deviations in the higher metallicity
regime.  This is indicative of the high sensitivity of the C4668 index to small variations in
$\alpha$-element abundance.

\begin{table}
\centering
\scriptsize
\caption{Columns 2 and 3 give the $\sigma_{rms}$ between [E/Fe]=0.3 TMK04 and TMK+KMT05 models (see
Section \ref{application}) in two different metallicity ranges.  In most cases the errors in our
enhancement application are well within our Lick calibration errors and, therefore, have little
effect on the derived parameters.}
\begin{tabular}{lccc}
\hline
Lick            &$\Delta_{TMK+KMT05}$   &$\Delta_{TMK+KMT05}$  \\
Index           &[Z/H]$\leq 0$  &[Z/H]$> 0$ \\
\hline
H$_{\delta A}$  (\AA)  &  0.068 &  0.221 \\
H$_{\delta F}$  (\AA)  &  0.024 &  0.074 \\
CN$_1$          (mag)  &  0.001 &  0.005 \\
CN$_2$          (mag)  &  0.001 &  0.005 \\
Ca4227          (\AA)  &  0.019 &  0.043 \\
G4300           (\AA)  &  0.059 &  0.081 \\
H$_{\gamma A}$  (\AA)  &  0.027 &  0.140 \\
H$_{\gamma F}$  (\AA)  &  0.014 &  0.023 \\
Fe4383          (\AA)  &  0.048 &  0.162 \\
Ca4455          (\AA)  &  0.002 &  0.004 \\
Fe4531          (\AA)  &  0.012 &  0.005 \\
C4668           (\AA)  &  0.133 &  0.444 \\
H$_\beta$       (\AA)  &  0.010 &  0.006 \\
Fe5015          (\AA)  &  0.027 &  0.043 \\
Mg$_1$          (mag)  &  0.003 &  0.007 \\
Mg$_2$          (mag)  &  0.002 &  0.005 \\
Mg~$b$          (\AA)  &  0.045 &  0.151 \\
Fe5270          (\AA)  &  0.018 &  0.011 \\
Fe5335          (\AA)  &  0.019 &  0.032 \\
Fe5406          (\AA)  &  0.016 &  0.011 \\
Fe5709          (\AA)  &  0.012 &  0.002 \\
Fe5782          (\AA)  &  0.012 &  0.005 \\
Na~D            (\AA)  &  0.007 &  0.014 \\
TiO$_1$         (mag)  &  0.000 &  0.000 \\
TiO$_2$         (mag)  &  0.001 &  0.001 \\
\hline
\end{tabular}
\label{conv_errors}
\end{table}

\clearpage
\clearpage

\onecolumn

\section{Index-Index Calibrations}
\label{ind_ind}
\begin{figure*}
\caption{A comparison of the common GCs between S05 and P02 using the Worthey \etal (1994) index
definitions for measurement and calibration.  Solid lines are a one-to-one correlation, with dashed
lines representing our adopted Lick/IDS conversion error as shown in Column 2 of Table
\ref{app_errors}.  Indices not included in fits to the SSP models (see Sect. \ref{fitting}) are
shown in grey.}
\centering
\includegraphics[scale=0.73,angle=0]{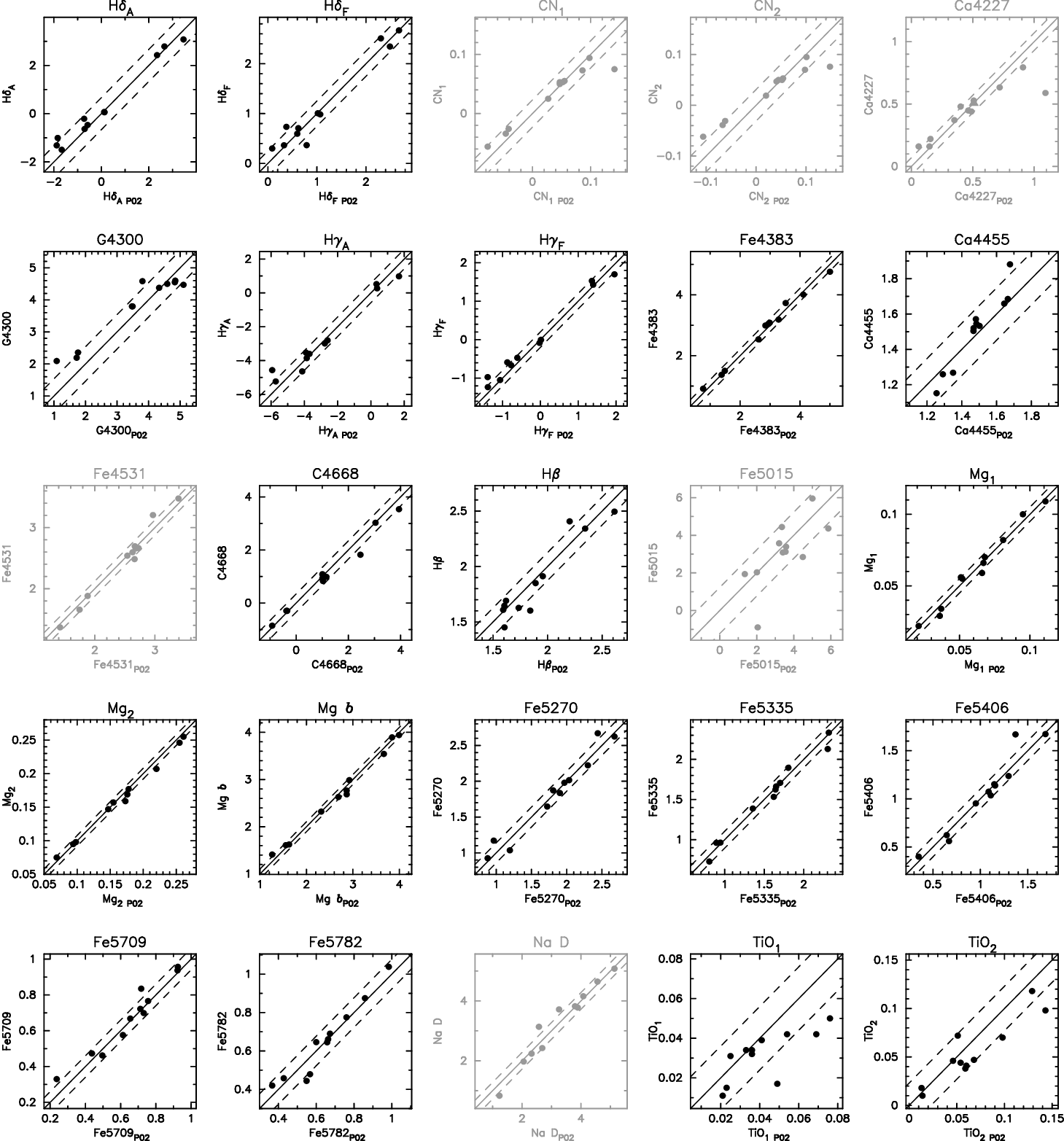}
\label{W94_ind}
\end{figure*}

\begin{figure*}
\caption{Identical to Fig. \ref{W94_ind}, but with indices measured and calibrated using the Trager
\etal (1998) index definitions.  Solid lines are a one-to-one correlation, with dashed lines
representing our adopted Lick/IDS conversion error as shown in Column 3 of Table \ref{app_errors}.}
\centering
\includegraphics[scale=0.73,angle=0]{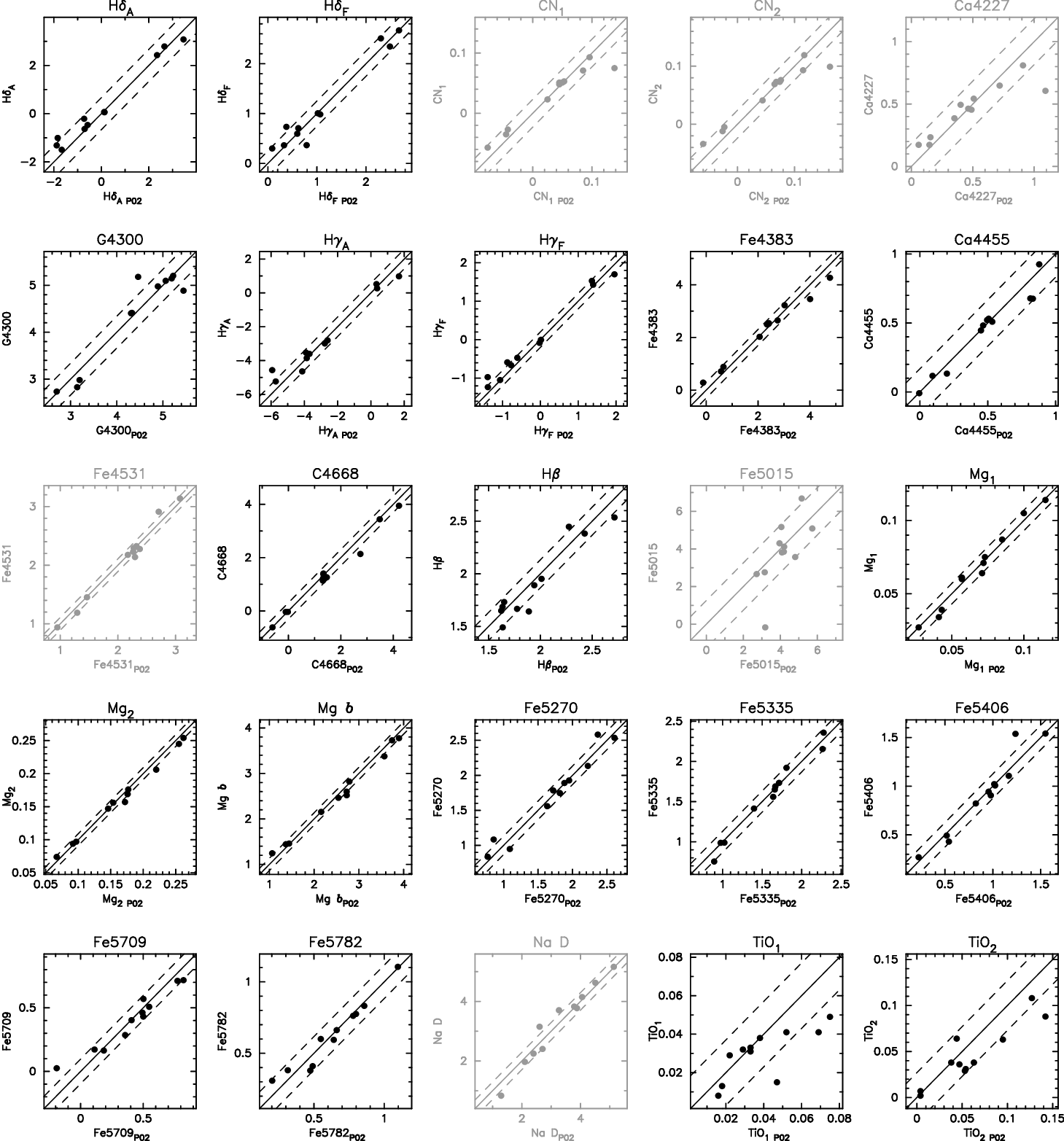}
\label{T98_ind}
\end{figure*}

\begin{figure*}
\caption{Identical to Fig. \ref{W94_ind}, but with indices measured using the Trager \etal (1998)
index definitions, but without any additional Lick calibration applied.  Solid lines are a
one-to-one correlation, with dashed lines representing our adopted Lick/IDS conversion error as
shown in Column 4 of Table \ref{app_errors}.}
\centering
\includegraphics[scale=0.73,angle=0]{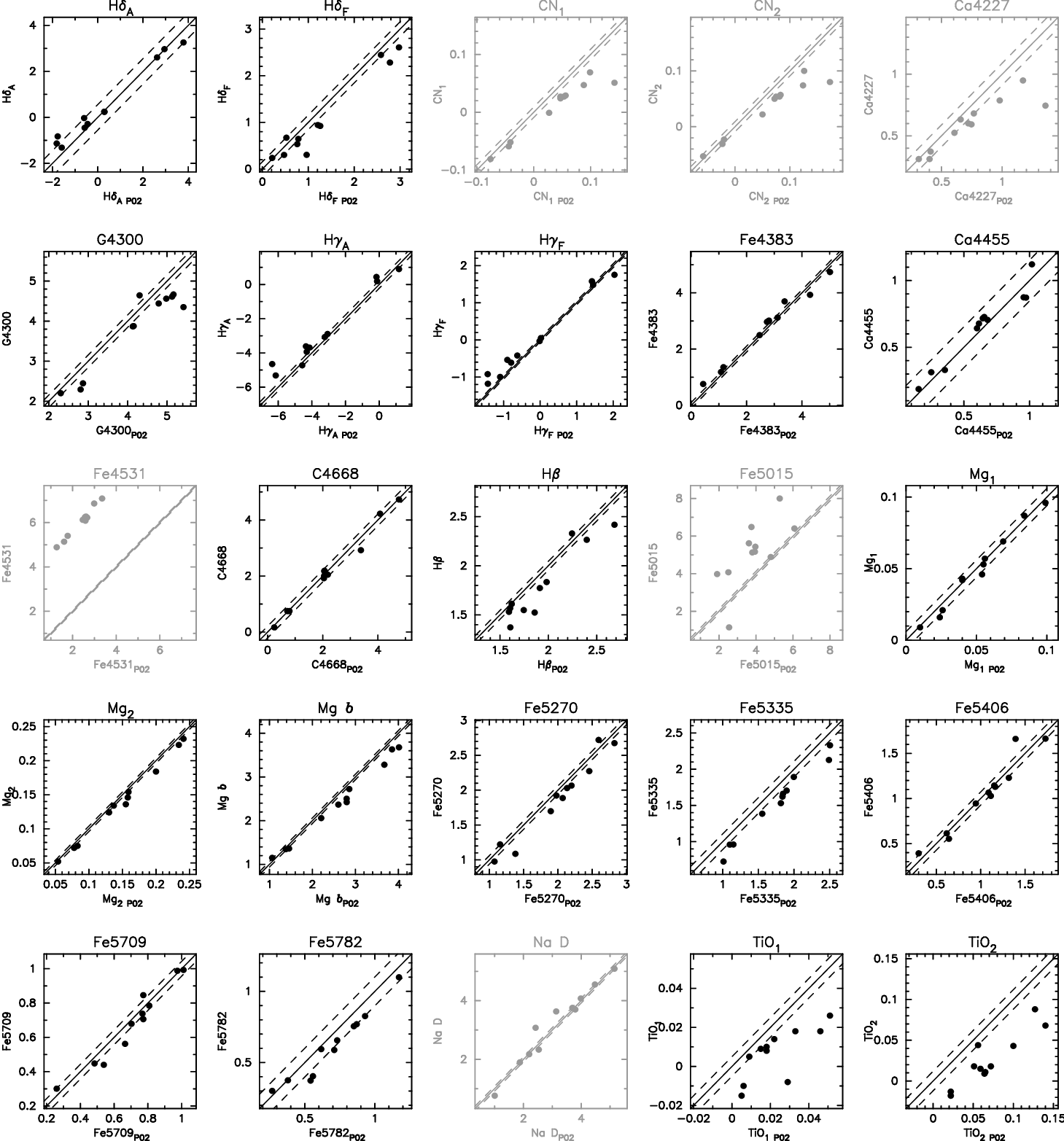}
\label{V06_ind}
\end{figure*}

Here we show the results of calibrating the S05 data to the P02 data.  Table \ref{app_errors} shows
the adopted error we have associated with these calibrations.  Figures \ref{W94_ind}, \ref{T98_ind}
and \ref{V06_ind} show index-index comparisons for the 11 GCs common between P02 and S05 for each of
the three different calibrations.

\begin{table}
\centering
\scriptsize
\caption{Summary of errors associated with the calibration of our GC data. Adopted Lick/IDS
calibration errors for S05 measured and calibrated using Worthey \etal (1994) and Trager \etal
(1998) index definitions are shown in Columns 2 and 3 respectively.\label{app_errors}}
\begin{tabular}{lccc}
\hline
Lick            &W94 Data     & T98 Data & T98 w/o Lick \\
Index           &$\sigma_{rms}$ &$\sigma_{rms}$ &$\sigma_{rms}$ \\
\hline
H$_{\delta A}$  (\AA)  & 0.665  & 0.665 & 0.553 \\
H$_{\delta F}$  (\AA)  & 0.252  & 0.252 & 0.156 \\
CN$_1$          (mag)  & 0.023  & 0.023 & 0.009 \\
CN$_2$          (mag)  & 0.031  & 0.024 & 0.009 \\
Ca4227          (\AA)  & 0.166  & 0.184 & 0.089 \\
G4300           (\AA)  & 0.537  & 0.339 & 0.155 \\
H$_{\gamma A}$  (\AA)  & 0.579  & 0.579 & 0.222 \\
H$_{\gamma F}$  (\AA)  & 0.200  & 0.200 & 0.052 \\
Fe4383          (\AA)  & 0.213  & 0.303 & 0.104 \\
Ca4455          (\AA)  & 0.148  & 0.168 & 0.153 \\
Fe4531          (\AA)  & 0.123  & 0.108 & 0.055 \\
C4668           (\AA)  & 0.343  & 0.305 & 0.209 \\
H$_\beta$       (\AA)  & 0.126  & 0.133 & 0.045 \\
Fe5015          (\AA)  & 1.228  & 1.258 & 0.128 \\
Mg$_1$          (mag)  & 0.005  & 0.007 & 0.006 \\
Mg$_2$          (mag)  & 0.008  & 0.009 & 0.004 \\
Mg~$b$          (\AA)  & 0.105  & 0.135 & 0.063 \\
Fe5270          (\AA)  & 0.139  & 0.132 & 0.061 \\
Fe5335          (\AA)  & 0.105  & 0.131 & 0.108 \\
Fe5406          (\AA)  & 0.102  & 0.126 & 0.073 \\
Fe5709          (\AA)  & 0.060  & 0.097 & 0.043 \\
Fe5782          (\AA)  & 0.077  & 0.117 & 0.101 \\
Na~D            (\AA)  & 0.279  & 0.292 & 0.081 \\
TiO$_1$         (mag)  & 0.016  & 0.017 & 0.005 \\
TiO$_2$         (mag)  & 0.023  & 0.027 & 0.012 \\
\hline
\end{tabular}
\end{table}

\section{Data calibrated using Trager \etal (1998) index definitions}
\label{trager}

As discussed in Section \ref{calibration}, P02 provide data measured and calibrated using both the
W94 and T98 index definitions, however in the text we use only the W94 calibrated data.  Here we
reproduce Figs. \ref{fe_z}--\ref{amr} using the P02 and S05 data measured and calibrated using T98
index definitions.

\vspace{5cm}
\begin{figure} 
\centering 
\includegraphics[scale=0.55,angle=-90]{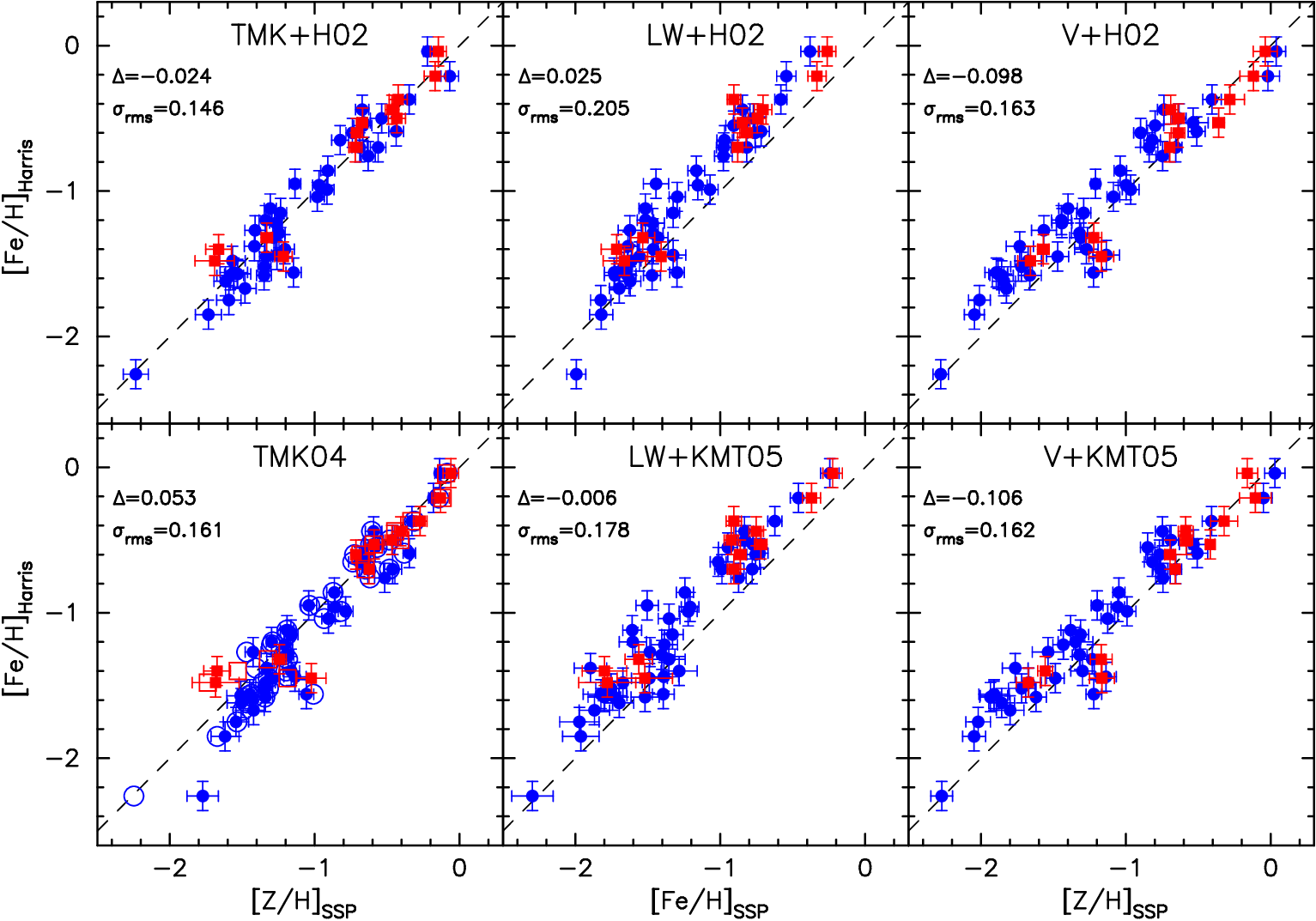} 
\caption{[Fe/H] from Harris \etal (1996) plotted against SSP derived metallicities. Numbers in the
upper left corner represent the mean offset from the one-to-one line (dashed) and the $\sigma_{rms}$
scatter about that offset, error bars signify a 1$\sigma$ deviation on our SSP fits and $\pm$
0.1\,dex for the Harris [Fe/H] values.} 
\label{T98_fe_z} 
\end{figure}

\begin{figure*} 
\centering 
\includegraphics[scale=0.55,angle=-90]{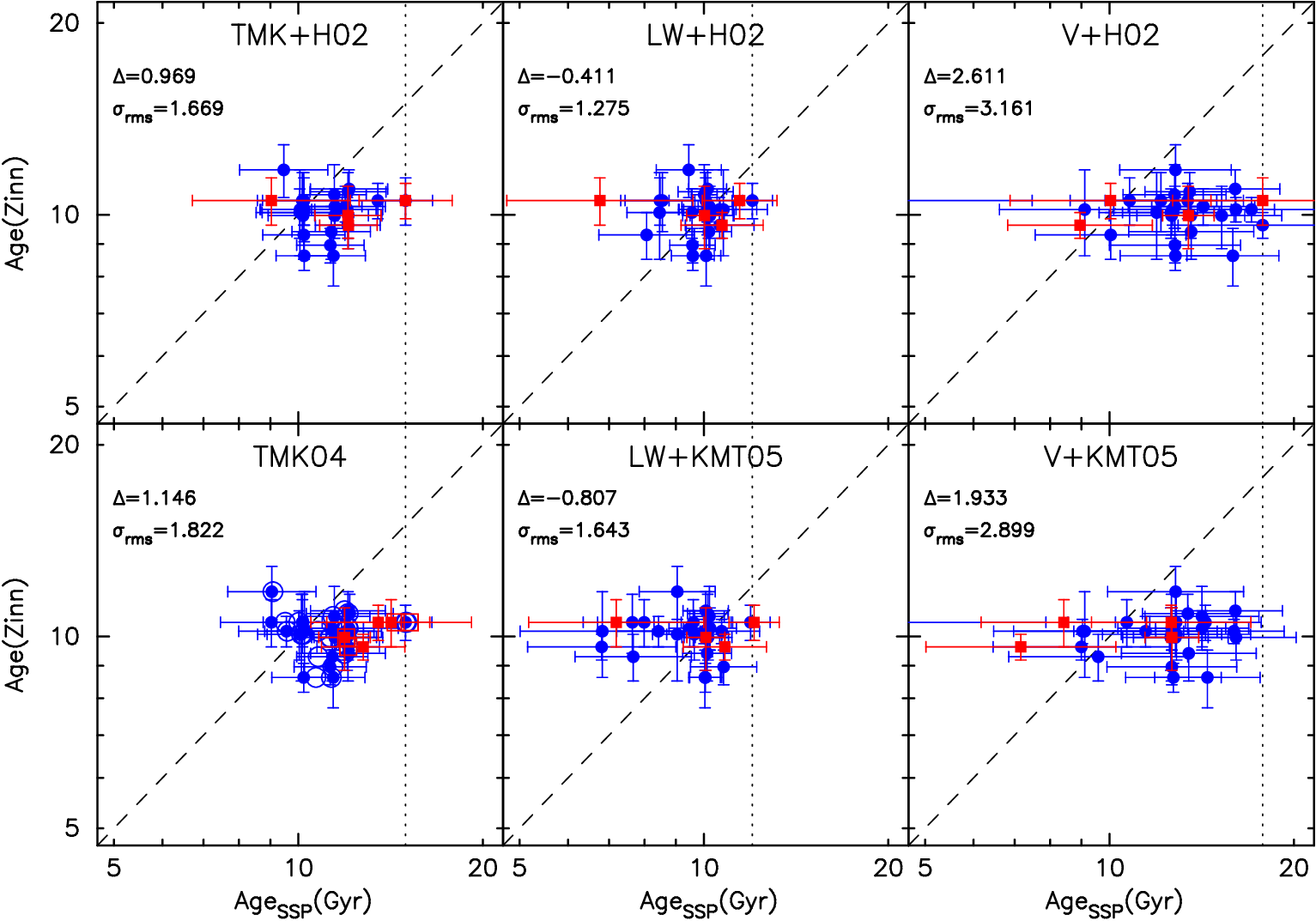}
\caption{Zinn \& West (1984) scale ages from De Angeli \etal (2005) plotted against SSP derived
ages, the dotted line in this case represents the oldest age in each model set.  Symbols represent
P02 (squares) and S05 (circles).  Numbers in the upper left corner represent the mean offset from
the one-to-one line (dashed) and the $\sigma_{rms}$ scatter about that offset, error bars signify a
1$\sigma$ deviation on our SSP fits.} 
\label{T98_age_age} 
\end{figure*}

\begin{figure*}
\centering
\includegraphics[scale=0.9,angle=-90]{figC3.eps}
\caption{SSP derived values of [E/Fe] plotted against high resolution element abundances from Pritzl
\etal (2005).  Models and symbols are the same as in Figure \ref{T98_fe_z}.  Error bars signify a
1$\sigma$ deviation on our SSP fits and $\pm$ 0.1\,dex for the high-resolution [$\alpha$/Fe]
values.} 
\label{T98_e_e}
\end{figure*}
  
\begin{figure*}
\centering
\includegraphics[scale=0.65,angle=-90]{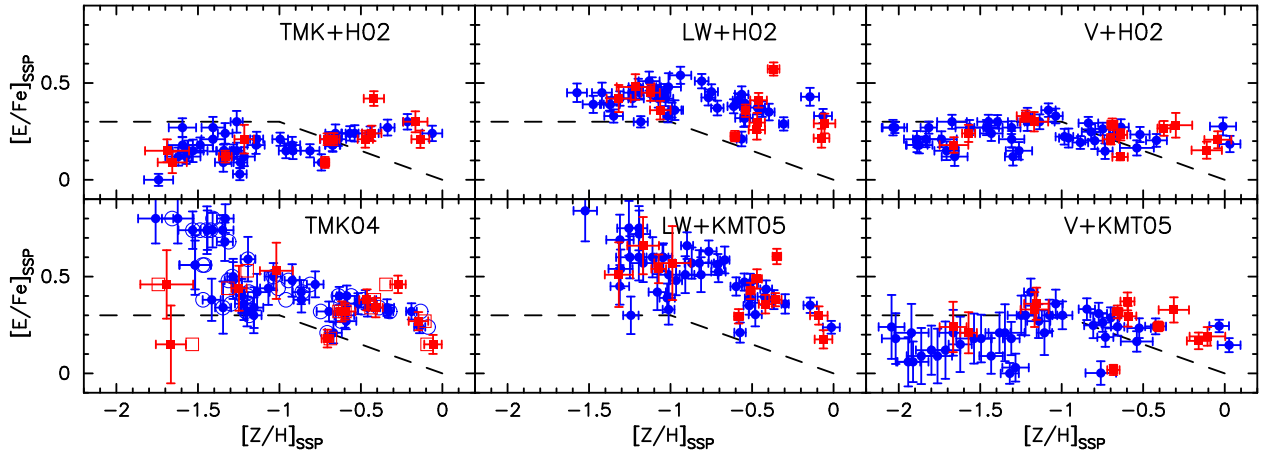}
\caption{Comparison of fitting results for TMK04, LW05 and V07 based models. Symbols are the same
as in previous figures.  The dashed line shows the assumed local abundance pattern in stars.  GCs
fit at the minimum SSP metallicities are not shown.} 
\label{T98_e_z}
\end{figure*}

\begin{figure*}
\centering
\includegraphics[scale=0.55,angle=-90]{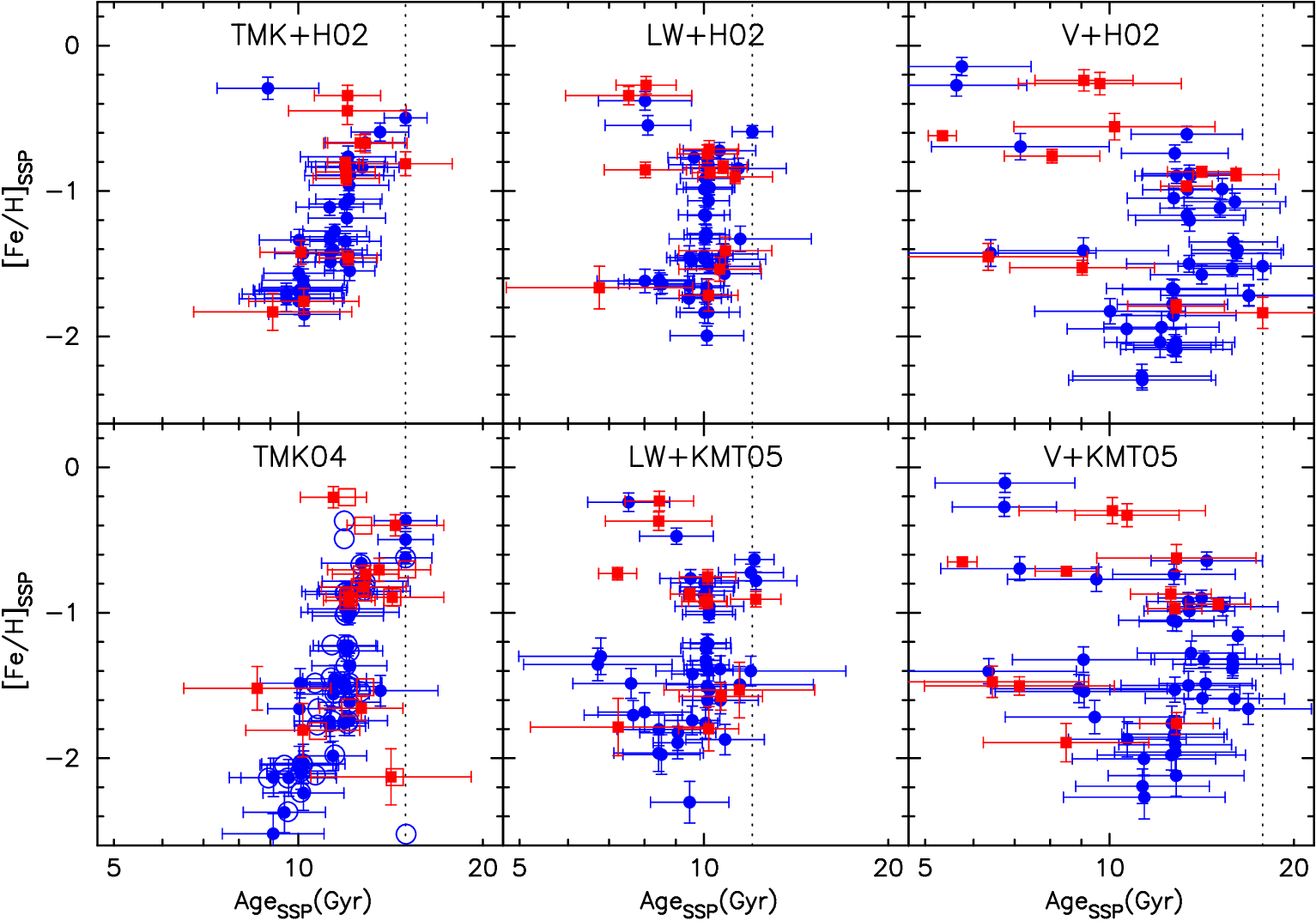}
\caption{[Fe/H] vs. age as derived from our SSP fitting.  Symbols and models are the same as in
previous figures.  The dotted line represents the maximum age for a particular SSP.} 
\label{T98_amr}
\end{figure*}

\clearpage
\clearpage

\end{appendix}
\end{document}